\documentclass[twocolumn,epjc3]{svjour3}  
\smartqed  % flush right qed marks, e.g. at end of proof
\RequirePackage{graphicx}
\journalname{Eur. Phys. J. C}

\usepackage{color}
\usepackage[makeroom]{cancel}
\usepackage[normalem]{ulem}

\usepackage[utf8]{inputenc}
\usepackage[T1]{fontenc}
\usepackage{array}
\usepackage{bm}
\usepackage{dsfont}
\usepackage{mathtools}
\usepackage{mathrsfs}
\usepackage[squaren,Gray]{SIunits}
\usepackage{array}
\usepackage{booktabs}
\usepackage{widetable}
\usepackage{widetext}
\usepackage{appendix}
\usepackage{bbold}

\usepackage{graphicx}

\definecolor{kkcolor}{rgb}{0,0.6,0.4}

\definecolor{cali_color}{rgb}{0,0.4470,0.7410}

\definecolor{pkcolor}{rgb}{0,0.1,0.7}

\begin{document}

%% Use the option review to obtain double line spacing
%% \documentclass[authoryear,preprint,review,12pt]{elsarticle}

%% Use the options 1p,twocolumn; 3p; 3p,twocolumn; 5p; or 5p,twocolumn
%% for a journal layout:
%% \documentclass[final,1p,times]{elsarticle}
%% \documentclass[final,1p,times,twocolumn]{elsarticle}
%% \documentclass[final,3p,times]{elsarticle}
%% \documentclass[final,3p,times,twocolumn]{elsarticle}
%% \documentclass[final,5p,times]{elsarticle}
%% \documentclass[final,5p,times,twocolumn]{elsarticle}

%% For including figures, graphicx.sty has been loaded in
%% elsarticle.cls. If you prefer to use the old commands
%% please give \usepackage{epsfig}

\newcommand{\subsubsubsection}[1]{\paragraph{#1}\mbox{}\\}
\setcounter{secnumdepth}{4}
\newcommand{\stat}{\textrm{stat}\,}
\newcommand{\sech}{\textrm{sech}\,}
\newcommand{\csch}{\textrm{csch}\,}
\newcommand{\kin}{\textrm{kin}\,}
\newcommand{\spin}{\textrm{spin}\,}
\newcommand{\bare}{\textrm{bare}\,}
\newcommand{\pole}{\textrm{pole}\,}
\newcommand{\ren}{\textrm{ren}\,}
\newcommand{\match}{\textrm{match}\,}
\newcommand{\HQET}{\textrm{HQET}\,}
\newcommand{\QCD}{\textrm{QCD}\,}
\newcommand{\msbar}{\overline{MS}\,}

\newcommand{\oo}{\mathcal{O}}
\newcommand{\lat}{\mathrm{lat}}
\newcommand{\free}{\mathrm{free}}
\newcommand{\massless}{\mathrm{massless}}
\newcommand{\cont}{\mathrm{cont}}
\newcommand{\MSb}{\overline{\mathrm{MS}}}
\newcommand{\MSt}{\widetilde{\mathrm{MS}}}

\newcommand{\latt}{\mathrm{latt}}
%% The lineno packages adds line numbers. Start line numbering with
%% \begin{linenumbers}, end it with \end{linenumbers}. Or switch it on
%% for the whole article with \linenumbers.
%% \usepackage{lineno}

%----------------------- temporary for editing  -----------------------------

\newcommand\pkout{\marginpar{\color{pkcolor}$\clubsuit$}\bgroup\markoverwith{\color{pkcolor}{\rule[0.4ex]{2pt}{0.8pt}}}\ULon}
\newcommand*{\pkcomment}[1]{{\color{pkcolor}\bf [PK: #1]}}
% for PKotko
\definecolor{pkotcolor}{rgb}{0.8,0,0}
\long\def\PKot#1{\marginpar{\color{pkotcolor}$\clubsuit$}{\color{pkotcolor} #1}}
\newcommand\pkotout{\marginpar{\color{pkotcolor}$\clubsuit$}\bgroup\markoverwith{\color{pkotcolor}{\rule[0.4ex]{2pt}{0.8pt}}}\ULon}
\newcommand*{\pkotcomment}[1]{{\color{pkotcolor}\bf [PKot: #1]}}




\title{On systematic effects in the numerical solutions of the JIMWLK equation%\thanksref{t1}
}


\author{Salvatore Cal\`i\thanksref{addr1,mit}
        \and
        Krzysztof Cichy\thanksref{addr2}
        \and
        Piotr Korcyl\thanksref{e1,addr1, addr3}
        \and
        Piotr Kotko\thanksref{addr4}
        \and
        Krzysztof Kutak\thanksref{addr5}
        \and
        Cyrille Marquet\thanksref{addr6}
}

\thankstext{e1}{e-mail: piotr.korcyl@uj.edu.pl}


\institute{Institute of Theoretical Physics, Jagiellonian University, ul. \L ojasiewicza 11, 30-348 Krak\'ow, Poland \label{addr1}
           \and
           Center for Theoretical Physics, Massachusetts Institute of Technology, Cambridge, MA 02139, USA
           \label{mit}
           \and
           Faculty of Physics, Adam Mickiewicz University, Uniwersytetu Pozna\'nskiego 2, 61-614 Pozna\'n, Poland \label{addr2}
           \and
           Institut f\"ur Theoretische Physik, Universit\"at Regensburg, 93040 Regensburg, Germany \label{addr3}
           \and
           AGH University of Science and Technology, Physics Faculty, ul. Mickiewicza 30, 30-059 Krakow, Poland \label{addr4}
           \and
           Institute of Nuclear Physics Polish Academy of Sciences, PL-31342 Krak\'{o}w, Poland \label{addr5}
           \and
           CPHT, CNRS, Ecole Polytechnique, Institut Polytechnique de Paris, 91128 Palaiseau, France \label{addr6}
}

\date{Received: date / Accepted: date}

\maketitle

\begin{abstract}
In the high energy limit of hadron collisions, the evolution of the gluon density in the longitudinal momentum fraction can be deduced from the Balitsky hierarchy of equations or, equivalently, from the nonlinear Jalilian-Marian-Iancu-McLerran-Weigert-Leonidov-Kovner (JIMWLK) equation. The solutions of the latter can be studied numerically by using its reformulation in terms of a Langevin equation. In this paper, we present a comprehensive study of systematic effects associated with the numerical framework, in particular the ones related to the inclusion of the running coupling. We consider three proposed ways in which the running of the coupling constant can be included: ``square root'' and ``noise'' prescriptions and the recent proposal by Hatta and Iancu.
We implement them both in position and momentum spaces and we investigate and quantify the differences in the resulting evolved gluon distributions. We find that the systematic differences associated with the implementation technicalities can be of a similar magnitude as differences in running coupling prescriptions in some cases, or much smaller in other cases.

\keywords{Heavy Ion Phenomenology \and QCD Phenomenology \and Langevin equation \and Numerical simulations}
\PACS{12.38.Lg \and 12.38.Mh \and 12.39.St}
\end{abstract}

\section{Introduction}

One of the well known predictions of perturbative Quantum Chromodynamics
(QCD) is a rapid growth of gluon distribution functions at a fixed resolution
scale $\mu$ with decreasing values of 
hadron longitudinal momentum
fraction $x$. 
For sufficiently small $x$, the growth is expected
to be tamed, because otherwise the unitarity bound would be violated.
This phenomenon -- the so-called gluon saturation -- is
one of the most interesting and still not fully resolved aspects of
QCD at high energies. For a comprehensive and pedagogical review of
these subjects, see, for instance, Ref.~\cite{Kovchegov:2012mbw}. 

Pictorially, in the saturation domain, the occupation number of gluons
inside a hadron is so large that they start to overlap. Consequently,
a new perturbative mechanism is possible, the gluon recombination,
which competes with the gluon splitting giving the aforementioned
growth of the distribution. At a fixed resolution scale $\mu$, the
growth due to the gluon splitting with decreasing $x$ is described
by the linear Balitsky-Fadin-Kuraev-Lipatov (BFKL) evolution equation
\cite{Fadin:1975cb,Kuraev:1977fs,Balitsky:1978ic}, while the inclusion
of the recombination effects leads to non-linear equations, the simplest
example being the Balitsky-Kovchegov (BK) equation \cite{Balitsky:1995ub,Kovchegov:1999yj}.

The BK equation is the mean field approximation to an infinite tower
of entangled equations, describing the evolution in $x$ of correlators
of Wilson line operators $\mathcal{U}\left(x_{T}\right)$, stretching
on the light-cone from minus to plus infinity, but positioned at a fixed
transverse point $x_{T}$ (with respect to the light-cone directions).
Such Wilson line operators appear naturally at high energies in the
so-called eikonal approximation, as phases picked up by energetic
colored parton fields traveling through the color field of the target
hadron.
The simplest two-point correlator of Wilson lines, the dipole, i.e.\ 
$\left\langle \mathcal{U}\left(x_{T}\right)\mathcal{U}^{\dagger}\left(y_{T}\right)\right\rangle $,
is directly related to the gluon distribution function unintegrated
over the transverse momentum, often called 
unintegrated dipole gluon density or transverse momentum dependent
(TMD) gluon density\footnote{Here, we will use the latter term, which should, however, not be confused with the TMD in the context of Collins-Soper-Sterman formalism \cite{Collins:1984kg,Collins:2011zzd}. } probed in inclusive processes such as deep inelastic scattering. 

In general, TMD gluon distributions turn out to be non-universal;
various processes enforce a different gauge link structure required
to maintain the gauge invariance of the operator defining the TMD
distribution \cite{Bomhof:2006dp}.
However, a TMD distribution for an arbitrary multiparticle process can
%\pkotout{ but any TMD distribution can }
be constructed from a restricted set of basis operators \cite{Bury:2018kvg}.
Correlators of four and more Wilson lines can be related to these
non-universal TMD gluon distributions -- the relation has been first
shown at leading power \cite{Dominguez:2010xd,Dominguez:2011wm,Marquet:2016cgx} and recently beyond leading twist \cite{Altinoluk:2019fui,Altinoluk:2019wyu,Boussarie:2020vzf}.

A systematic approach to the evolution in energy (or $x$) of the
correlators of Wilson lines is provided by the Color Glass Condensate
(CGC) theory 
(see e.g. \cite{Gelis:2010nm}). 
In the CGC, the Wilson lines are treated as functionals of the target (typically nucleus) random gauge field.
The averaging of the Wilson line
operators is defined by means of the functional weight, which is typically
taken
to be the Gaussian functional
at some initial scale $x_{0}$, 
according to the McLerran-Venugopalan (MV) model  \cite{McLerran:1993ni,McLerran:1993ka}.
The decrease of the scale modifies the weight functional according
to the Balitsky-Jalilian-Marian-Iancu-McLerran-Weigert-Leonidov-Kovner
(B-JIMWLK) equations \cite{Balitsky:1995ub,JalilianMarian:1997jx,JalilianMarian:1997gr,JalilianMarian:1997dw,Kovner:2000pt,Kovner:1999bj,Weigert:2000gi,Iancu:2000hn,Ferreiro:2001qy}, available now also at NLL accuracy \cite{PhysRevD.88.111501,Kovner:2013ona,Lublinsky:2016meo}.
Since the evolution Hamiltonian involves the Wilson lines themselves, the set
of equations for correlators is not closed, i.e.\ each step in the
evolution will generate more and more complicated correlators. Due
to this last fact, the aforementioned BK equation is phenomenologically
very important, because in the mean field approximation it allows
to approximate any correlator by dipoles. However, the precise description
of less inclusive processes, where the non-universa\-lity of TMD gluon
distributions is strong,  requires full solutions of the B-JIMWLK equations.

In particular, recent phenomenology studies of forward jet production processes \cite{vanHameren:2016ftb,Kotko:2017oxg,Albacete:2018ruq,vanHameren:2019ysa,Bury:2020ndc}  within a framework that directly uses various  TMD gluon distributions in the small-$x$ limit -- the so-called small-$x$ improved TMD factorization (ITMD) \cite{Kotko:2015ura,Altinoluk:2019fui} --  utilizes the Gaussian approximation and large-$N_c$ approximation to calculate the TMD distributions. 
As shown in \cite{Marquet:2016cgx,Marquet:2017xwy}, they can also be calculated from the full B-JIMWLK equation, without approximations, although at a fixed coupling constant.

It is possible to study the solutions of the JIMWLK equation numerically by rewriting it in terms of 
a more practical Langevin equation \cite{Rummukainen:2003ns}. The first application of such numerical solutions to phenomenology was described in  \cite{Mantysaari:2018zdd} and used the so-called ``square root'' prescription for the inclusion of effects of the running of the strong coupling constant. We devote this work to investigate the impact of the various systematic effects inherent to the numerical framework, such as the differences which arise from performing particular steps of the calculation in position or momentum space. The Langevin equation itself is defined only in position space. However, intermediate or additional steps, such as the computation of correlation function of Wilson lines, can be implemented both in position and momentum spaces. There are two reasons why the Fourier transform introduces unwanted systematic effects. 
On the one hand, the Fourier transforms mix short-distance discretization effects with large-distance finite volume effects. 
It is difficult to disentangle them and one needs to perform both the continuum and the infinite volume extrapolations to remove them. On the other hand, the Fourier transforms are in many cases performed in the color group instead of in the
corresponding algebra and therefore many of the theorems known in Fourier analysis do not hold. We show numerically that under some circumstances the two ways of estimation, through position or momentum space, are not equivalent and we quantify these effects.
In particular, we pay special attention to the various prescriptions for the running coupling constant effects and we try to quantify the associated differences.

Our results can be used as guidelines to estimate systematic uncertainties on the phenomenological parameters of the initial condition in the MV model. The recent phenomenological fit to the $F_2$ HERA data provided first estimates of these parameters. They are, however, dependent on different implementation choices of the numerical framework used to solve the JIMWLK evolution equation. We argue that some care needs to be taken when these parameters are compared with experimental determinations. 

The paper is composed as follows. We start in Section~\ref{sec. framework} with a brief summary of the main features of the numerical framework. In Section~\ref{sec. running coupling}, we describe in more detail existing implementations of 
the running coupling in the JIMWLK equation. In subsequent sections, we discuss the systematics associated to different stages of the computation. Some subtleties even at the stage of computing the correlation functions are mentioned in Section~\ref{sec. correlation function}. In Section~\ref{sec. initial condition}, we summarize the dependence of the initial gluon distribution on the algorithmic parameters, rederiving some of the results of \cite{Rummukainen:2003ns} and \cite{Marquet:2016cgx} for completeness. Further, in Section~\ref{sec. evolution}, we describe the systematics involved in the implementation of the evolution in rapidity. Finally, in Section~\ref{sec. running coupling systematics}, we discuss the differences induced by different implementations of the running coupling in the JIMWLK equation. We present our conclusions in Section~\ref{sec. conclusions}.

Throughout this paper, we try to keep an explicit notation of all expressions. We find that in the literature many details of the implementation are implicitly assumed, but never precisely stated. Although some of our expressions seem to be trivial, their aim is to provide the Reader with expressions that correspond exactly to what was implemented in the computer code
used to perform the computations. The open-source code is available as a git repository under \\
\begin{center}
\verb[https://bitbucket.org/piotrekkorcyl/jimwlk[
\end{center}
together with the documentation. The code is fully parallelized and multi-threaded and can be run on multicore CPU-based clusters. General information about the code can be found in Ref.~\cite{Korcyl:2020orf}.

%===========================================================================================

\section{Description of the numerical framework}
\label{sec. framework}

In this section, we describe in detail the basic steps of the calculation. We describe the implementation of the MV model for the initial condition and present the computation of the dipole correlation function. Then, we proceed with the description of the evolution equation, providing details of the implementation entirely in position space or using the JIMWLK kernel in momentum space. The inclusion of the running coupling effects is described in the next section.

\subsection{Initial condition and construction of Wilson lines}

The basic object in our calculation is the
straight infinite Wilson line
along the light cone with fixed transverse position $\mathbf{x}$. In the discretized setting, it is constructed as the product of $N_y$ elementary Wilson links,
\begin{equation}
\label{eq. wilson line}
U^{ab}(\mathbf{x}) = \prod_{k=1}^{N_y} U^{ab}_k(\mathbf{x}) \, .
\end{equation}
The infinitesimal Wilson link variables can be obtained from the gauge potentials $A_k^{ab}$
\begin{equation}
    U_k^{ab}(\mathbf{x}) = \exp\left( -i g A_k^{ab}(\mathbf{x}) \right) = 
    \exp\left(-i \frac{g \rho^{ab}_k(\mathbf{x})}{\nabla^2 + m^2} \right) \, ,
\label{eq. exponentiation}
\end{equation}
which are expressed in terms of color source fields $\rho^{ab}_k$ and imposed to be the solutions of appropriate classical Yang-Mills equations, in accordance with the MV model.
In the approximation of stationary, classical fields, 
these equations reduce to a single Poisson equation,
\begin{equation}
    (\nabla^2 + m^2)  A^{ab}_k(\mathbf{x}) =  \rho^{ab}_k(\mathbf{x}) \,,
\label{eq. poisson}
\end{equation}
where the artificial mass $m$ regulates a possible zero eigenvalue of the Laplace operator. 
We solve Eq.~\eqref{eq. poisson} on the square lattice with periodic boundary conditions. The Laplace operator is diagonal in momentum space, so the simplest way of obtaining the solution is to perform the Fourier transform element-wise on the $\rho$ matrix, solve the corresponding algebraic equation in momentum space and return to position space,
\begin{multline}
\frac{\rho^{ab}_k(\mathbf{x})}{\nabla^2 + m^2} =\\=
\frac{a^2}{L^2} \sum_{\mathbf{z} \in \tilde{\Lambda}} \sum_{\mathbf{k} \in \Lambda} 
\frac{  e^{i\mathbf{k}(\mathbf{x} - \mathbf{z})}\rho^{ab}_k(\mathbf{z})}{ - \frac{4}{a^2}\left[\sin^2\left(\frac{k_x a}{2}\right) + \sin^2\left(\frac{k_y a}{2}\right)\right] + m^2} \,,
\label{eq. poisson details}
\end{multline}
where the two sums correspond to the two Fourier transforms and $\Lambda$ ($\tilde{\Lambda}$) denote the sets of all discrete lattice momenta (positions), see \ref{sec. appendix} for details. The lattice has size $L_x\times L_y$ and we take $L_x=L_y\equiv L$. $a$ is the dimensionful lattice spacing. The dimensionless lattice momentum 
\begin{equation}
    \hat{k}_{x,y} = \frac{2}{a}\sin\left(\frac{k_{x,y} a}{2}\right), 
\end{equation}
and its square
\begin{equation}
    \hat{\mathbf{k}}^2 = \frac{4}{a^2}\left\{ \sin^2\left(\frac{k_x a}{2}\right) + \sin^2\left(\frac{k_y a}{2}\right) \right\}
    \label{eq. lattice momentum}
\end{equation}
arise in Eq. \eqref{eq. poisson} as a direct consequence of using a symmetrized finite difference instead of the continuum deri\-vative. 

For the remaining problem of generating initial color sources $\rho$, we follow the MV model with a discretized transverse $x\mathrm{-}y$ plane. For a given $\mathbf{x}=(x,y)$ on the transverse plane, $N_y$ color matrices in the rapidity direction represent successive, random gluon radiation. Such implementation follows the procedure described in Ref.~\cite{Lappi:2007ku}. Alternative implementations were discussed in Ref.~\cite{Rummukainen:2003ns}.

Technically, on each site $\mathbf{x}$ of each plane $k$, $k=1,\dots,N_y$, we construct a random matrix from the SU(3) algebra by 
\begin{equation}
g \rho(\mathbf{x})_k = g \rho(\mathbf{x})_k^a \lambda^a \,,
\end{equation}
where the color sources $\rho(\mathbf{x})_k^a$ are normally distributed and generated by the Box-Muller method. For their standard deviation, we have
\begin{equation}
\langle g \rho(\mathbf{x})_k^a  g \rho(\mathbf{y})_l^b \rangle = \delta^{ab} \delta^{kl} \delta(\mathbf{x} - \mathbf{y}) \frac{g^4\mu^2}{N_y} \,.
\end{equation}
Above, $g$, $N_y$ and $\mu$ are input parameters of the MV model. As can be seen from the above 
expression, the relevant combination of MV parameters is $g^2 \mu$. It is therefore sufficient to set $g=1$ and vary $\mu$. Summarizing, the Wilson line is given by 
\begin{equation}
 U^{ab}(\mathbf{x}) = \prod_{k=1}^{N_y} \exp\left(-i \frac{ \rho^{ab}_k(\mathbf{x})}{\nabla^2 + m^2}\right) \,.
\end{equation}
On the lattice, the dimensionless (input) scale parameter is $a\mu$ and hence, a fixed physical setting is obtained when multiplying it with the lattice size $L/a$. Thus, in our numerical study, we take a constant value of $g^2 \mu L = 30.72$, as used in Refs.~\cite{Lappi:2012vw,Marquet:2016cgx}.

\subsection{Two-point correlation function}
\label{sec. correlation function}

Once we have constructed all Wilson lines, we can evaluate their correlation functions. At first sight, this seems to be a trivial task. However, depending on how the estimation of the correlation function is implemented, i.e. whether one stays entirely in position space or one uses Fourier acceleration and performs parts of the calculation in momentum space, the results differ significantly on a single realization of the initial condition. This is due to the ambiguity of the Fourier transform on the group manifold. The discrepancy disappears when a sufficiently large statistical sample is used for the evaluation. Below, we summarize both approaches and the numerical results are shown in Fig.~\ref{fig. correlation function}.
\begin{figure}
\begin{center}
\includegraphics[width=0.5\textwidth]{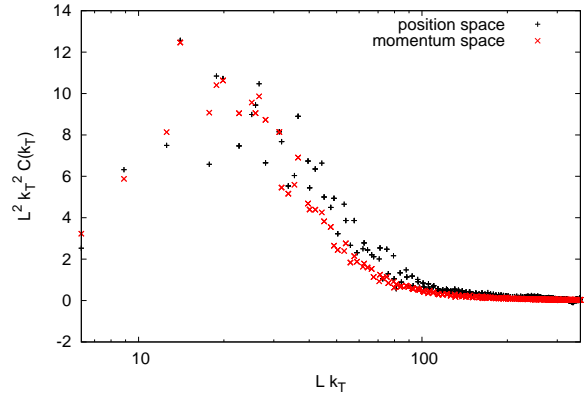}
\includegraphics[width=0.5\textwidth]{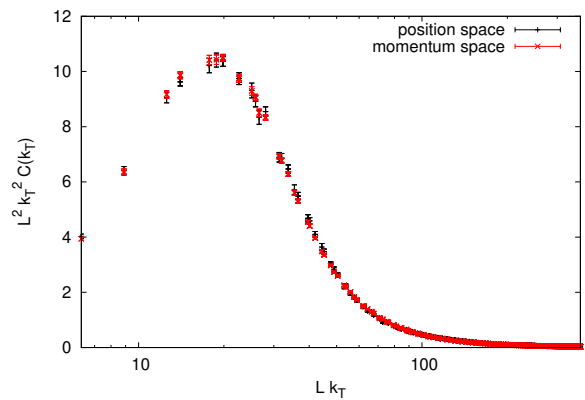}
\caption{Top: Comparison of the correlation function estimated for a single realization of the initial condition. Bottom: comparison of the correlation functions estimated in momentum and position spaces for a statistics of 100 initial condition realizations. \label{fig. correlation function}} 
\end{center}
\end{figure}
We stay in the fundamental representation and define $C(\mathbf{x})$ following Ref.~\cite{Lappi:2007ku},
\begin{equation}
\label{eq. correlation function position}
C(\mathbf{x} - \mathbf{y}) = \langle \textrm{tr}\, U^{\dagger}(\mathbf{x}) U(\mathbf{y}) \rangle,
\end{equation}
where the average $\langle \cdot \rangle$ is taken over different statistical realizations of Wilson lines.
We enforce that $C(\boldsymbol{\delta})$ is an even function of $\boldsymbol{\delta} = \mathbf{x} - \mathbf{y}$, 
\begin{equation}
C(\delta_x,\delta_y) = C(|\delta_x|, |\delta_y|) \,.
\end{equation}
We average over symmetric distances
\begin{equation}
    C(\delta_x,\delta_y) = \frac{1}{2} \big\{ C(\delta_x,\delta_y) 
    + C(\delta_y,\delta_x)  \big\}
\end{equation}
and
\begin{multline}
    C(\delta_x,\delta_y) = \frac{1}{4} \big\{ C(\delta_x,\delta_y) 
    + C(L - \delta_x,\delta_y) +\\+ C(\delta_x,L - \delta_y)
    + C(L - \delta_x,L - \delta_y) \big\},
\end{multline}
where $\delta_x \ge 0$ and $\delta_y \ge 0$.
Finally, we use the Fourier transform of complex-valued function $C$ with periodic boundary conditions to obtain
\begin{equation}
\label{eq. correlation function mom}
\tilde{C}(\mathbf{k}) = \frac{a}{L} \sum_{\mathbf{x}-\mathbf{y} \in \tilde{\Lambda}} e^{i \mathbf{k}( \mathbf{x}-\mathbf{y})} C(\mathbf{x}-\mathbf{y}),
\end{equation}
which, due to the applied symmetrization, yields a strictly real function $\tilde{C}(\mathbf{k})$. 

A much faster implementation uses Wilson lines in momentum space. The starting point is the Fourier transform of Wilson lines,
\begin{equation}
\tilde{U}^{ab}(\mathbf{k}) = \frac{a}{L} \sum_{\mathbf{x} \in \tilde{\Lambda}} e^{i \mathbf{k} \mathbf{x}} U^{ab}(\mathbf{x}) \,.
\end{equation}
This Fourier transform is understood in terms of matrix elements of the matrices $U$, i.e.\ the above equation involves 9 independent complex number Fourier transforms.
Obviously, in general, $\tilde{U}(\mathbf{k})$ does not belong to the SU(3) group anymore, 
\begin{equation}
\tilde{U}^{\dagger}(\mathbf{k})\tilde{U}(\mathbf{k}) = \frac{a^2}{L^2} \sum_{\mathbf{x},\mathbf{y} \in \tilde{\Lambda}} e^{i \mathbf{k} (\mathbf{x} - \mathbf{y})} U^{\dagger}(\mathbf{x}) U(\mathbf{y}) \ne 1 \, .
\end{equation}
The advantage is that using $\tilde{U}(\mathbf{k})$ we can recover the correlation function of Eq.~(\ref{eq. correlation function mom}) by only local multiplications, 
\begin{multline}
\label{eq. correlation function momentum}
\tilde{C}(\mathbf{k}) = \langle \textrm{tr} \, \tilde{U}^{\dagger}(\mathbf{k}) \tilde{U}(\mathbf{k}) \rangle =\\= \frac{a^2}{L^2}  \sum_{\mathbf{x},\mathbf{y} \in \tilde{\Lambda}} e^{i \mathbf{k} (\mathbf{x} - \mathbf{y})} \langle \textrm{tr} \, U^{\dagger}(\mathbf{x}) U(\mathbf{y}) \rangle \,, 
\end{multline}
where we have used the fact that $\langle \textrm{tr} \, U^{\dagger}(\mathbf{x}) U(\mathbf{y}) \rangle$ depends only on the distance $\mathbf{x}-\mathbf{y} = \boldsymbol{\delta}$. Numerical results for both implementations are shown in Fig.~\ref{fig. correlation function}. In the top panel, we present the results for both approaches for a single configuration of Wilson lines. Clearly, the results differ significantly, especially for small values of the transverse momentum vector. On the contrary, after averaging over 100 realizations of the color charges, both methods give compatible results (bottom panel), the momentum space approach being statistically more precise. In the following, we use the momentum space approach to evaluate all correlation functions.

Below, we present results for the rescaled dimensionless gluon distribution, $(L/a)^2 \ \hat{k}^2 \ \tilde{C}(L k_T)$, where 
$k_T$ is the norm of the two-dimensional vector $\mathbf{k}$ and where we use the lattice momentum, Eq.~\eqref{eq. lattice momentum}, for $\hat{k}^2$, as previously suggested in the literature. For clarity of our message,
we do not consider in this work other correlation functions, in particular those containing derivatives of Wilson lines, such as the ones presented in Ref.~\cite{Marquet:2016cgx}. They share all the systematic uncertainties with the basic gluon distribution analyzed below, their additional systematic uncertainties being solely related to the discretization effects of the derivative.

\subsection{Evolution in rapidity}

In this section, we describe the details of the implementation of the JIMWLK equation reformulated as 
a Langevin equation \cite{Rummukainen:2003ns}. We use a numerically more economical formulation, where the Wilson line is multiplied by evolution kernels from the left and right side \cite{Lappi:2012vw}. The evolution equation in rapidity $s$ with a step of size $\delta s$ reads

\begin{widetext}
\begin{equation}
\label{eq. main}
U(\mathbf{x},s+\delta s) = \exp\left( -\sqrt{\delta s} \sum_{\mathbf{y}} U(\mathbf{y},s) \left(\mathbf{K}(\mathbf{x}-\mathbf{y}) \cdot \boldsymbol{\xi}(\mathbf{y}) \right) U^{\dagger}(\mathbf{y},s) \right) \, U(\mathbf{x},s) \,  
\exp \left( \sqrt{\delta s} \sum_{\mathbf{y}} \mathbf{K}(\mathbf{x}-\mathbf{y}) \cdot \boldsymbol{\xi}(\mathbf{y}) \right),
\end{equation}
\end{widetext}
\noindent where $\mathbf{K}(\mathbf{x})$ is a kernel function and $\boldsymbol{\xi}(\mathbf{x})$ are random vectors, both to be discussed below.
To simplify the notation, we denote the arguments of the right and left exponentials by $A$ and $B$, 
\begin{equation}
    \label{eq. A}
    A^{ab} = \sum_{\mathbf{y}} \mathbf{K}(\mathbf{x}-\mathbf{y}) \cdot \boldsymbol{\xi}^{ab}(\mathbf{y})
\end{equation}
and
\begin{equation}
    \label{eq. B}
    B^{ab} = \sum_{\mathbf{y}} U^{ac}(\mathbf{y},s) \left(\mathbf{K}(\mathbf{x}-\mathbf{y}) \cdot \boldsymbol{\xi}^{cd}(\mathbf{y}) \right) U^{\dagger,db}(\mathbf{y},s),
\end{equation}
so that the Langevin equation becomes
\begin{equation}
U(\mathbf{x},s + \delta s) = \exp(-\sqrt{\delta s}B) \, U(\mathbf{x},s) \, \exp(\sqrt{\delta s}A) \,.
\label{eq. evolution equation short}
\end{equation}
Again, the construction of $A$ and $B$ can be performed in position or momentum spaces. Below, we describe both approaches and the corresponding numerical results are discussed in Section \ref{sec. evolution}.

\subsubsection{Construction of $A$ and $B$}
\label{sec. construction of A and B}

In order to use Fourier acceleration, we need several quantities in momentum space. We define 
\begin{eqnarray}
\mathbf{\tilde{K}}(\mathbf{k}) = \frac{a}{L} \sum_{\mathbf{x} \in \tilde{\Lambda}} e^{i \mathbf{k} \mathbf{x}} \mathbf{K}(\mathbf{x}), \\
\boldsymbol{\tilde{\xi}}(\mathbf{k}) = \frac{a}{L} \sum_{\mathbf{x} \in \tilde{\Lambda}} e^{i \mathbf{k} \mathbf{x}} \boldsymbol{\xi}(\mathbf{x}).
\end{eqnarray}
Then,
\begin{equation}
A(\mathbf{x}) =  \frac{a}{L} \sum_{\mathbf{k} \in \Lambda} e^{-i \mathbf{k} \mathbf{x}}   \tilde{\mathbf{K}}(\mathbf{k})  \tilde{\boldsymbol{\xi}}(\mathbf{k}) \,.
\end{equation}
In order to repeat a similar construction for the $B$ matrix, we start by defining a matrix $\mathbf{\mathcal{U}}$
\begin{equation}
\mathbf{\mathcal{U}}(\mathbf{y}) = U(\mathbf{y}) \boldsymbol{\xi}(\mathbf{y}) U^{\dagger}(\mathbf{y})
\,,
\end{equation}
which we subsequently transform to momentum space
\begin{equation}
\tilde{\mathcal{U}}(\mathbf{k}) = \frac{a}{L} \sum_{\mathbf{x} \in \tilde{\Lambda}} e^{i \mathbf{k} \mathbf{x}}   \mathcal{U}(\mathbf{x}) \,.
\end{equation}

\noindent Then, we have
\begin{equation}
B(x) = \frac{a}{L} \sum_{\mathbf{k} \in \Lambda} e^{-i \mathbf{k} \mathbf{x}}  \tilde{\mathbf{\mathcal{U}}}(\mathbf{k})
\tilde{\mathbf{K}}(\mathbf{k})  \,.
\end{equation}

All Fourier transforms above are understood ele\-ment-wise. Because of that and because of the different discretization errors in the JIMWLK kernel, both approaches, in position and in momentum space, are not equivalent. The evolution equation in momentum space reads

\begin{widetext}
\begin{equation}
\label{eq. main}
U(\mathbf{x},s+\delta s) = \exp\left( -\sqrt{\delta s} 
\frac{a}{L} \sum_{\mathbf{k} \in \Lambda} e^{-i \mathbf{k} \mathbf{x}}  \tilde{\mathbf{\mathcal{U}}}(\mathbf{k})
\tilde{\mathbf{K}}(\mathbf{k}) 
\right) \, U(\mathbf{x},s) \,  
\exp \left( \sqrt{\delta s} 
\frac{a}{L} \sum_{\mathbf{k}} e^{-i \mathbf{k} \mathbf{x}}
\tilde{\mathbf{K}}(\mathbf{k}) \tilde{\mathbf{\xi}(\mathbf{k})}
\right) \,,
\end{equation}
\end{widetext}

Further numerical results are discussed in Section \ref{sec. systematics}.

\subsubsection{Random vectors}
In both approaches, we start by generating random vectors $\boldsymbol{\xi}$ valued in the Lie algebra (with generators $\lambda^a$) on each site of the lattice,
\begin{equation}
\boldsymbol{\xi}(\mathbf{x}) = (\xi_x(\mathbf{x}), \xi_y(\mathbf{x})) = (\xi_x^a(\mathbf{x}) \lambda^a, \xi_y^b(\mathbf{x}) \lambda^b) \,,
\end{equation}
from a normal distribution with unit width. The vectors $\boldsymbol{\xi}(\mathbf{x})$ are uncorrelated in $\mathbf{x}$, therefore
\begin{equation}
\label{eq. noise}
\langle \xi^a_i(\mathbf{x})\, \xi^b_j(\mathbf{y}) \rangle = \delta^{ab} \delta_{ij}\, \delta(\mathbf{x} - \mathbf{y}) \,. 
\end{equation}
For the momentum space formulation, we also need the Fourier-transformed random vectors $\boldsymbol{\xi}(\mathbf{p})$, which satisfy the analogous constraint
\begin{equation}
\langle \xi^a_i(\mathbf{p})\, \xi^b_j(\mathbf{q}) \rangle = \delta^{ab} \delta_{ij}\, \delta(\mathbf{p} - \mathbf{q}) \,, 
\end{equation}
which states that the noise vectors are also uncorrelated in momentum space. 

\subsubsection{Kernel function}

We now turn to the kernel function that enters the rapidity evolution equation and discuss 
the possible discretizations thereof in position space and in momentum space. The original JIMWLK kernel is defined in continuum position space as
\begin{equation}
    \mathbf{K}(\mathbf{x}) = \frac{\mathbf{x}}{x^2} \,.
    \label{eq kernel}
\end{equation}
When implemented on a lattice, several symmetries of this kernel are broken and hence, various discretizations are possible and lead to different systematic effects. In particular, due to the slow, power-law decay of the kernel at large distances, finite volume effects must be carefully studied. We start our discussion with the position space kernel and we show how to construct the $A$ and $B$ matrices. Then, we present an analogous presentation for the momentum space kernel.

\subsubsubsection{Position space kernel}

On a discrete lattice, the kernel in Eq.~\eqref{eq kernel} assumes the following form,
\begin{equation}
\label{eq:Kdiscrete}
\mathbf{K}(\mathbf{n})=\frac{\mathbf{\bar{n}}}{\bar{n}^2},
\end{equation}
where $\mathbf{n}=(n_x,n_y)$ is a vector of integers, i.e.\ $n_i\in(-L/a, L/a)$, and the numbers $\bar{n}_i$ ($\bar{n}^2)$ are in the chosen discretizations:  
\begin{itemize}
\item naive with a discontinuity:
\begin{eqnarray}
\bar{n}_i&=& 
\begin{cases}
    n_i-L/a & \text{if } n_i\geq L/2a,\\
    n_i & \text{if } -L/2a\leq n_i < L/2a, \\ 
    n_i+L/a & \text{if } n_i<-L/2a,
\end{cases} \\   
\bar{n}^2&=&\bar{n}_x^2+\bar{n}_y^2
\label{eq. position space kernel}
\end{eqnarray}

\item regularized with the sine function:
\begin{eqnarray}
\bar{n}_i&=&\frac{L}{2\pi a} \sin\left(\frac{2\pi a n_i}{L}\right),\\
\bar{n}^2&=&\left(\frac{L}{\pi a}\right)^2 \Big[ \sin^2\left(\frac{\pi a n_x}{L}\right) + \sin^2\left(\frac{\pi a n_y}{L}\right)\Big] \,.
\label{eq. position space kernel lattice}
\end{eqnarray}
\end{itemize}
The numerator/denominator of the discrete kernel are plotted in the top/bottom panel of Fig.~\ref{plot kernel} for both discretizations. By construction, they agree for small separations $\mathbf{n}$, whereas the largest discrepancy appears for separations close to the half of the lattice extent, which for the demonstrated situation happens at $|x/a|=16$.
\begin{figure}[t!]
\begin{center}
\includegraphics[width=0.5\textwidth]{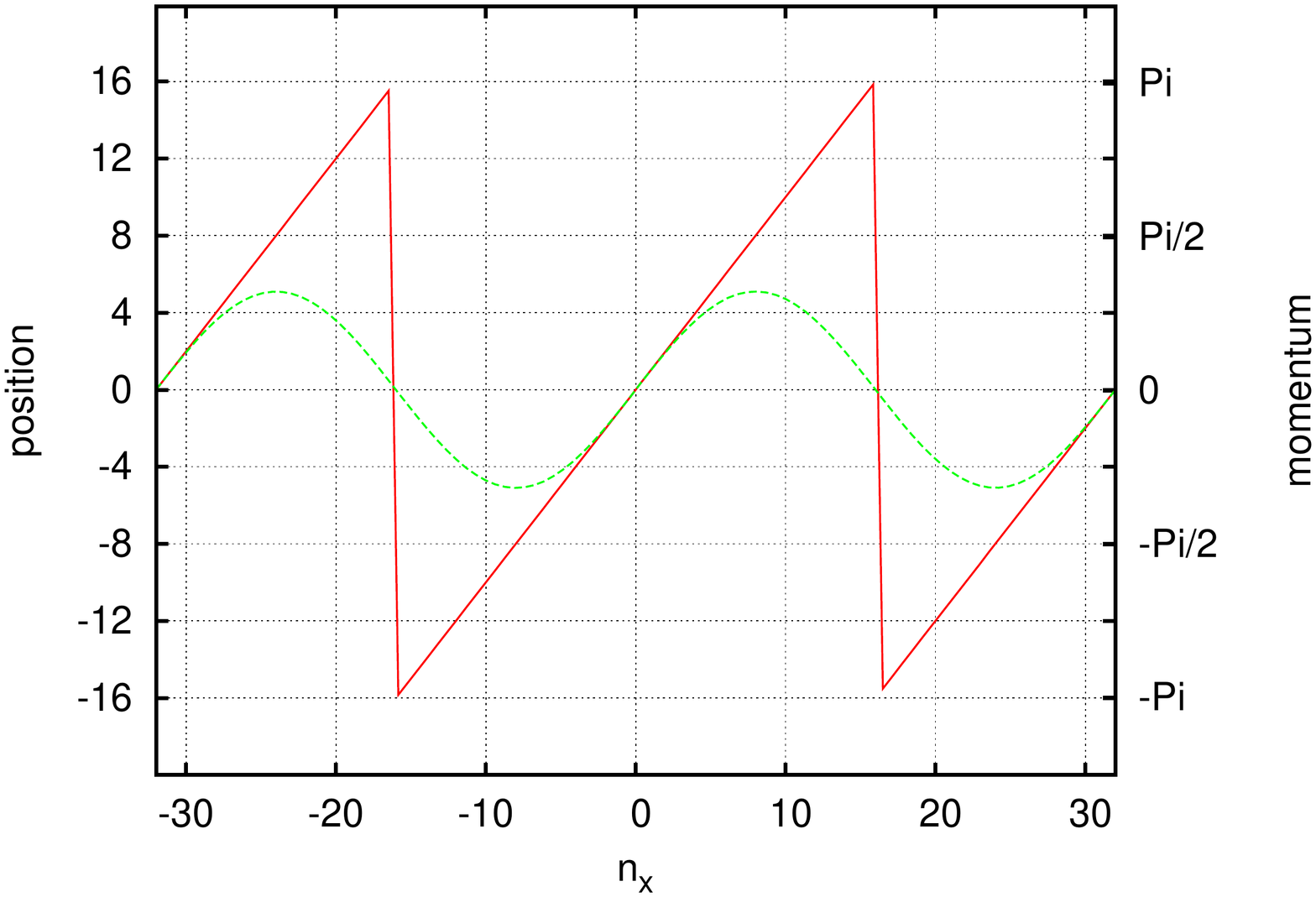}
\includegraphics[width=0.5\textwidth]{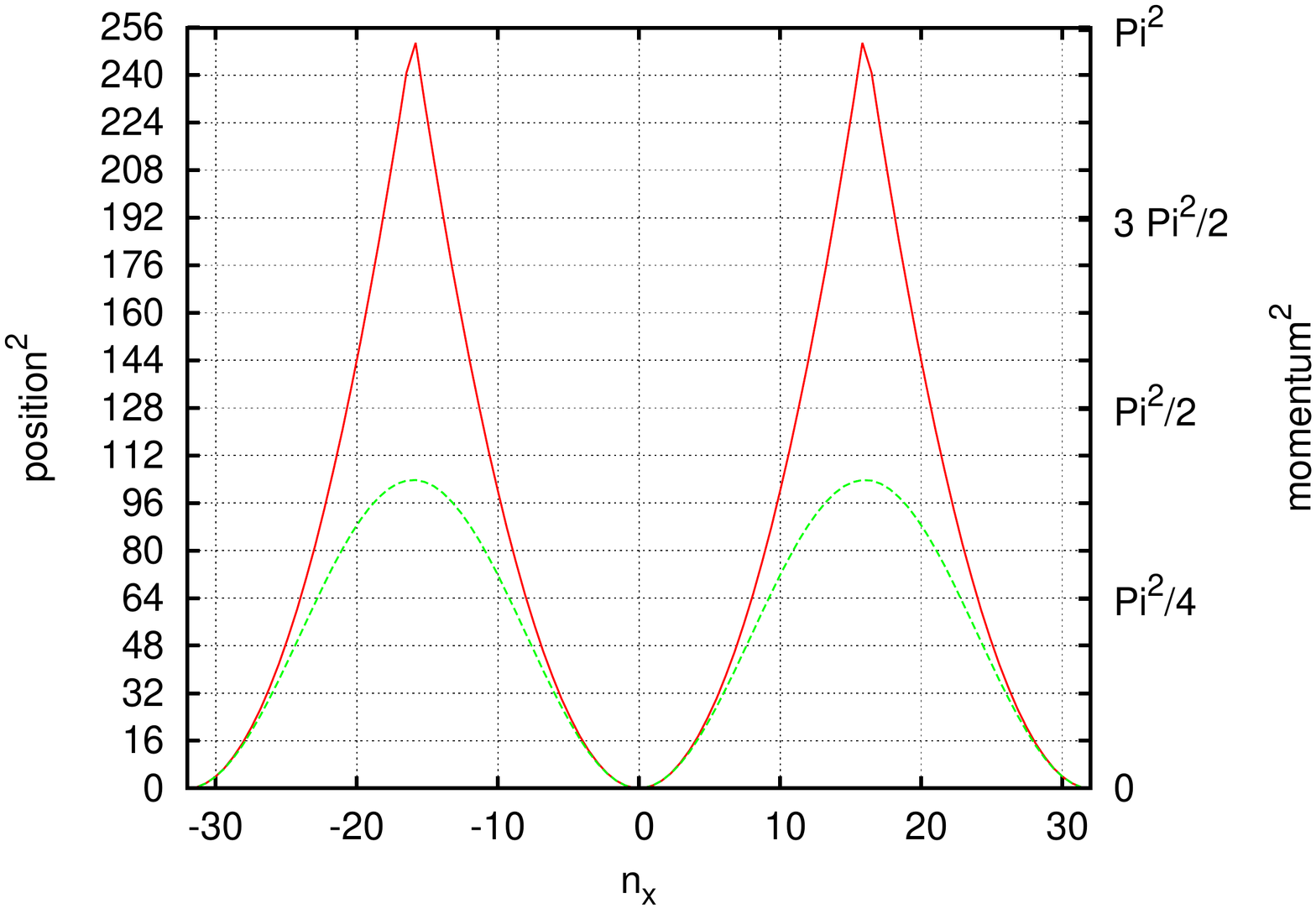}
\vspace{-0.75cm}
\caption{Top panel: periodic $x$-component of the distance vector $\mathbf{x}$ (left vertical axis) and the momentum vector $\mathbf{k}$ (right vertical axis), as appears in the numerator of the JIMWLK kernel. The naive/sine discretization is shown in red/green. Bottom panel: analogously for the squares of the distance and momentum vectors, corresponding to the denominator of the JIMWLK kernel. The lattice size is $L/a=32$. \label{plot kernel}}
\end{center}
\end{figure}

\subsubsubsection{Momentum space kernel}

In order to be able to compare position and momentum space results, we need to make sure that the relation between the JIMWLK kernel in position and momentum space is well established. As both sides of the equation have the same dimensions, we can write it in terms of dimensionless, still continuous, variables,
\begin{equation}
\int \frac{d^2 \mathbf{n}}{2\pi} e^{- i \mathbf{k} \mathbf{n}}  \frac{n_i}{n^2} = -2 \pi i \frac{k_i}{k^2} \,.
\label{eq. kernel relation}
\end{equation}
We have discussed possible discretizations of the integrand for the position space kernel. Mimicking the continuum relation, discrete Fourier transform should produce the equivalent JIMWLK kernel in momentum space. Anticipating the result, we can discretize the right hand side of the continuum relation, Eq.~\eqref{eq. kernel relation}, following one of the two ways,
\begin{itemize}
\item keeping the naive lattice momenta,
\begin{equation}
\bar{k}_i= 2 \pi \frac{a n_i}{L}.
\label{eq. momentum space kernel}
\end{equation}

\item or, alternatively, as is commonly done using the lattice momenta $\hat{k}$ and $\bar{k}$,
\begin{eqnarray}
\bar{k}_i&=& \sin\left(\frac{2\pi a n_i}{L}\right),\\
\bar{k}^2&=& 4 \Big[ \sin^2\left(\frac{\pi a n_x}{L}\right) + \sin^2\left(\frac{\pi a n_y}{L}\right) \Big] \,.
\label{eq. momentum space kernel lattice}
\end{eqnarray}
\end{itemize}

The comparison of both definitions in momentum space is shown
again in Fig.~\ref{plot kernel}. Obviously, both discretizations lead to the same behavior as in position space, up to a rescaling of the vertical axis.

\begin{figure*}[h]
\begin{center}
\includegraphics[width=0.75\textwidth]{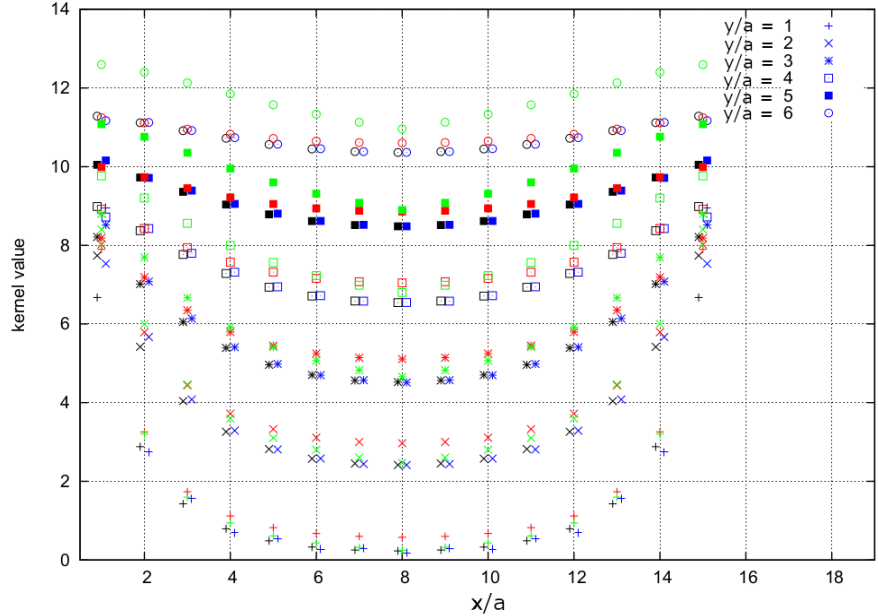}
\caption{Comparison of Fourier transformed position space JIMWLK kernels with their momentum definitions as in Eq.~\eqref{eq. fourier transform of kernel}. The horizontal axis shows $n_x$, whereas different values of $n_y$ are shown with different symbols and are shifted vertically by two units. The blue and black data points show the Fourier transform; the blue correspond to the naive discretization of the position space kernel, whereas the black to the $\sin$ regularized kernel. Surprisingly, both data sets agree quite well. The red data points show the corresponding momentum kernel with the usual $\bar{k}/\hat{k}^2$ definition, whereas the green data points are calculated using the simple lattice momenta. \label{plot kernel comparison}}
\end{center}
\end{figure*}
We can now test the discretized versions of Eq.~\eqref{eq. kernel relation}. Note that we should not \emph{a priori} expect that this equation holds in the discrete version. This is due to the fact that the Fourier transform propagates discretization effects and finite volume effects of the position space kernel to all values of the momentum $k$ and \emph{vice versa}. To simplify the picture, we look at a single component of the kernel, i.e.\ the $x$-component,
\begin{equation}
\int \frac{d n_x d n_y}{2\pi} e^{- i (k_x n_x + k_y n_y)}  \frac{\mathcal{D}^{\textrm{pos}}(n_x)}{\mathcal{D}^{\textrm{pos}}(n^2)} \approx -2 \pi i \frac{\mathcal{D}^{\textrm{mom}}(k_x)}{\mathcal{D}^{\textrm{mom}}(k^2)} \,,
\label{eq. fourier transform of kernel}
\end{equation}
where $\mathcal{D}^{\textrm{pos}}$ and $\mathcal{D}^{\textrm{mom}}$ denote one of the possible discretizations of the quantity in brackets, i.e.\ either the position or the momentum vector. In principle, there may not be a single region where Eq.~\eqref{eq. fourier transform of kernel} holds with a reasonable degree of accuracy. As a test (shown in Fig. \ref{plot kernel comparison}), we plug both definitions of the kernel in position space, perform the discretized Fourier transform numerically and compare the result with the two possible variants of the momentum space kernel. Note that we keep the usual dimensionless positions and momenta in the Fourier phase, and stick to the continuum notation in this sketchy equation. The result shown in Fig.~\ref{plot kernel comparison} was obtained for a lattice of size $16 \times 16$. The horizontal axis shows $x$, whereas different values of $y$ are shown with different symbols and are shifted vertically by two units. First, start with the left hand side of Eq.~\eqref{eq. fourier transform of kernel}, i.e. the Fourier transform of the position space kernel. This is shown by the blue and black data points; the blue correspond to the naive discretization of the position space kernel, whereas the black to the $\sin$-regularized kernel. Surprisingly, both data sets agree quite well for nearly all values of $x$ and also $y$. The red data points show the momentum kernel with the usual $\bar{k}/\hat{k}^2$ definition, whereas the green data points are calculated using the simple lattice momenta. 
If Eq.~\eqref{eq. fourier transform of kernel} held, the blue and black data points should lie on top of the green and red. The discrepancy signals that the linear momentum discretization fails to reproduce the discretized JIMWLK kernel in the bulk of the lattice: green data is significantly above all others. On the contrary, notice that the Fourier transforms of both position space kernels (blue and black data) agree reasonably well with the momentum kernel defined using trigonometric regularization. The momentum kernel defined through lattice momenta provides good agreement only for small momenta. This is reasonable, as the numerator of the kernel is linear in the lattice momentum, hence as seen in Fig.~\ref{plot kernel}, the naive and trigonometric momenta agree up to $\approx 6$, then the boundary effects introduce large deviations.

We conclude that both forms of the position space JIMWLK kernel and the trigonometric momentum kernel provide a consistent picture. This conclusion will be further enhanced when we discuss the gluon distribution after evolution in rapidity: the momentum linear kernel introduces large finite size effects and distorts the shape of the final distribution.

\section{Running coupling}
\label{sec. running coupling}

\subsection{``Square root'' prescription}

This prescription was proposed in Refs.~\cite{PhysRevD.75.014001,Kovchegov_2007} and discussed by
Rummukainen and Weigert in Ref.~\cite{Rummukainen:2003ns}, where they introduced the running coupling in the Lan\-ge\-vin formulation of the JIMWLK equation following \cite{MUELLER2002331} and \cite{TRIANTAFYLLOPOULOS2003293}. 
The running coupling effects are accounted for with the coupling at the scale given by the size of the parent dipole $(x-y)^2$, i.e. we introduce a one-loop running
\begin{equation}
\label{eq. alpha_s}
    \alpha_s \rightarrow \alpha_s(1/(x-y)^2) = \frac{4 \pi}{\beta_0 \ln \frac{1}{(x-y)^2 \Lambda^2_{\textrm{QCD}}}} \, ,
\end{equation}
with $\beta_0 = (11 N_c - 2 N_f)/3$ for $N_c$ colors and $N_f$ flavors and $\Lambda_{\textrm{QCD}}$ being the QCD scale parameter.
The running coupling in the Langevin equation (\ref{eq. main}) is hidden in the rapidity factor
\begin{equation}
s = \frac{\alpha_s}{\pi^2}y, \qquad y = \ln \frac{x_0}{x_2} \,,
\end{equation}
with $x_0$ ($x_2$) the initial (final) Bjorken-$x$ of the evolution.
In the ``square root'' prescription, Eq.~\eqref{eq. main} becomes
\begin{widetext}
\begin{multline}
U(\mathbf{x},s + \delta s) = \exp\Big[- \frac{\sqrt{\delta y}}{\pi} \sum_{\mathbf{y}} U(\mathbf{y},s) \Big( \sqrt{\alpha_s(|\mathbf{x}-\mathbf{y}|)} \mathbf{K}(\mathbf{x}-\mathbf{y}) \boldsymbol{\xi}(\mathbf{y}) \Big) U^{\dagger}(\mathbf{y},s)\Big] \ \times \\ \times U(\mathbf{x},s) 
 \exp\Big[\frac{\sqrt{\delta y}}{\pi} \sum_{\mathbf{y}} \sqrt{\alpha_s(|\mathbf{x}-\mathbf{y}|)} \mathbf{K}(\mathbf{x}-\mathbf{y}) \boldsymbol{\xi}(\mathbf{y})\Big],
\end{multline}
\end{widetext}
\noindent where we have separated
\begin{equation}
    \frac{\sqrt{\alpha_s \delta y}}{\pi} \rightarrow \frac{\sqrt{\delta y}}{\pi} \sqrt{\alpha_s(|\mathbf{x}-\mathbf{y}|)} \, .
\end{equation}
Depending on the way in which we have implemented the JIMWLK kernel, either in position or in momentum space, the running coupling has to be included accordingly. This question was addressed in Ref. \cite{Lappi:2012vw}, where the momentum prescription was provided.
We therefore make use of the following definitions,
\begin{equation}
\alpha_s(\mathbf{k}) = \frac{4 \pi}{\beta_0 \ln \big\{\big[\big(\frac{\mu_0^2}{\Lambda^2_{\textrm{QCD}}} \big)^{\frac{1}{c}} + \big(\frac{\mathbf{k}^2}{\Lambda^2_{\textrm{QCD}}} \big)^{\frac{1}{c}} \big]^{c} \big\} }
\label{eq. alpha_s in momentum}
\end{equation}
or
\begin{equation}
\alpha_s(\mathbf{r}) = \frac{4 \pi}{\beta_0 \ln \big\{\big[\big(\frac{\mu_0^2}{\Lambda^2_{\textrm{QCD}}} \big)^{\frac{1}{c}} + \big(\frac{4e^{-2 \gamma_E}}{ \mathbf{r}^2 \Lambda^2_{\textrm{QCD}}} \big)^{\frac{1}{c}} \big]^{c} \big\} } \, ,
\end{equation}
where the parameter $c$ is used to freeze the running for small momenta and large distances. As in Ref.~\cite{Lappi:2012vw} we take the numerical values of $L \mu_0 = 15$ and $L \Lambda_{\textrm{QCD}} = 6$. These parameters are part of the model and so their precise values should be fixed by comparison with experimental data, e.g.\ through a fit to the DIS data.\\

\subsection{``Noise'' prescription}

An alternative definition of the running coupling was proposed by Lappi and M\"antysaari \cite{Lappi:2012vw,Lappi:2014wya}. The running coupling can be implemented as a
modification of the properties of the noise vectors in the Langevin equation. This has a different
physical motivation, as the scale of the running coupling is in this case provided by the momentum of the emitted gluon. 
This scale is then argued in Refs.~\cite{Lappi:2012vw,Lappi:2014wya} to correspond to the smallest of the three
relevant  dipole  sizes  (the  “parent”  and  two  “daughter”
dipoles) \cite{Lappi:2012vw}. 
Hence, instead of Eq.~(\ref{eq. main}) we use
\begin{widetext}
\begin{equation}
U(\mathbf{x}, s + \delta s) = \exp\Big[- \sqrt{\delta s} \sum_{\mathbf{y}} U(\mathbf{y},s) \Big(\mathbf{K}(\mathbf{x}-\mathbf{y}) \boldsymbol{\eta}(\mathbf{y}) \Big) U^{\dagger}(\mathbf{y},s)\Big] \
 U(\mathbf{x},s) \ \exp\Big[\sqrt{\delta s} \sum_{\mathbf{y}}\mathbf{K}(\mathbf{x}-\mathbf{y}) \boldsymbol{\eta}(\mathbf{y})\Big] \,,
\end{equation}
\end{widetext}
\noindent where now
\begin{multline}
\langle \eta^{a,i}(\mathbf{x}) \eta^{b,j}(\mathbf{y}) \rangle = \delta^{ab} \delta^{ij} \int \frac{d^2\mathbf{k}}{(2\pi)^2} e^{i \mathbf{k}(\mathbf{x} - \mathbf{y})} \alpha_s(\mathbf{k}) = \\
= \delta^{ab} \delta^{ij} \hat{\alpha}_{\mathbf{x} - \mathbf{y}}
\label{eq. noise coupling constant}
\end{multline}
instead of 
the uncorrelated, Gaussian random variables in position space $\xi^{a,i}(\mathbf{x})$. The noise $\boldsymbol{\eta}(\mathbf{x})$ becomes correlated in both momentum and position spaces.

\subsubsection{Momentum space}

The implementation in momentum space is straightforward. The correlation is diagonal and thus, for each $\mathbf{p}$ one generates uncorrelated Gaussian variable with variance $\sigma = \sqrt{\alpha_s(\mathbf{p})}$,
\begin{equation}
\langle \eta^{a,i}(\mathbf{p}) \eta^{b,j}(\mathbf{q}) \rangle = \delta^{a,b} \delta^{i,j} \delta(\mathbf{q} - \mathbf{p}) \alpha_s(\mathbf{p}) \, .
\end{equation}

\subsubsection{Position space}

In position space, we need to construct a nontrivial correlation between any two lattice sites,
\begin{equation}
\langle \eta^{a,i}(\mathbf{x}) \eta^{b,j}(\mathbf{y}) \rangle = \delta^{a,b} \delta^{i,j} \hat{\alpha}_{\mathbf{x} - \mathbf{y}} \, .
\end{equation}
As noted in Ref.~\cite{Lappi:2012vw}, the coupling constant $\tilde{\alpha}_{\mathbf{x} - \mathbf{y}}$ cannot be interpreted as being evaluated at the scale $\mathbf{x} - \mathbf{y}$, but rather it is the Fourier transform of the coupling constant in momentum space, as in Eq.~\eqref{eq. noise coupling constant}. Hence, we construct the desired correlation matrix in position space, 
\begin{equation}
\Sigma(n,m) = \hat{\alpha}_s(n-m) \equiv \int \frac{d^2\mathbf{k}}{(2\pi)^2} e^{i \mathbf{k}(\mathbf{x} - \mathbf{y})} \alpha_s(\mathbf{k}) \,.
\end{equation}
$\Sigma$ is by construction symmetric and positive-definite.
Thus, in the next step, we find the Cholesky decomposition of $\Sigma$, i.e.\ a matrix $L$ such that
\begin{equation}
L L^{T} = \Sigma \,.
\end{equation}
Eventually, we generate a Gaussian vector of uncorrelated random variables, $\xi$, which we 
afterwards transform according to
\begin{equation}
\eta(n) = \sum_m L(n,m) \xi(m) \,.
\end{equation}
$\eta(n)$ has the desired correlation,
\begin{equation}
\langle \eta(n) \eta(m) \rangle = \hat{\alpha}_{\mathbf{n} - \mathbf{m}} \,.
\end{equation}
The correlation matrix $\Sigma$ is color and spin blind, so that we can use it for all color and spin components
\begin{equation}
\langle \eta^{a,i}(n) \eta^{b,j}(m) \rangle = \delta^{ij} \delta^{ab} \hat{\alpha}_{\mathbf{n} - \mathbf{m}} \,,
\end{equation}
as postulated by Lappi and M\"antysaari. A robust comparison of different implementations is shown in Fig.~\ref{fig. evolution comparison}. 
We note that for both running coupling prescriptions  the position space and the momentum space evolutions lead to consistent results within our statistical uncertainties. Although, for this small rapidity no significant deviation between the ``square root'' and ``noise'' prescription can be seen, for large rapidities these two prescriptions give significantly different saturation scales, as we will show below.

\begin{figure}
\begin{center}
\includegraphics[width=0.5\textwidth]{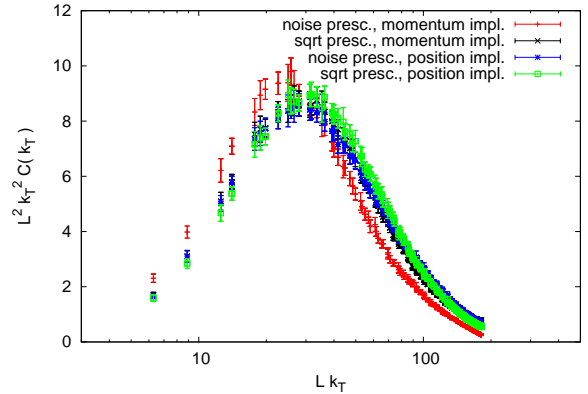}
\caption{Rescaled dimensiolness gluon distributions at rapidity $s=0.05$ on a $L/a=64$ lattice. We compare 'square root' and noise prescriptions for the running coupling, both for position-space evolution and momentum-space evolution. All cases are evolved from the same initial condition. 
\label{fig. evolution comparison}} 
\end{center}
\end{figure}

\subsection{Hatta-Iancu prescription}
In Ref.~\cite{Hatta:2016ujq}, the Authors provide a formulation of the JIMWLK equation with collinear re\-sum\-mation,\ \ which accounts for DGLAP kind of logarithms suppressing anti-collinear pole of the kernel. In particular, they re-in\-ves\-ti\-ga\-te the relation of momentum space expression for the running of $\alpha_s$ to small dipole prescription discussed in \cite{Lappi:2012vw,Lappi:2014wya}. According to Hatta and Iancu, the smallest dipole prescription corresponds to the dependence of $\alpha_s$ on virtuality and not the transverse momentum. 
We proceed with implementation directly in coordinate space,
\begin{equation}
    \alpha_s = \alpha_s( \min\{ |\mathbf{x}-\mathbf{y}|, r \} ) \, ,
\end{equation}
where $r$ is the size of the projectile.
We implement this prescription with the position space evolution only, by fixing the distance $r$ at which the Wilson line correlator is evaluated and performing the entire evolution for that given distance. This adds an additional $L$ factor to the scaling of computations, but in this way we avoid using Fourier transforms, as it is not clear how the Fourier transform would mix the contributions from distances different than $r$ where the coupling constant would be evaluated at a wrong scale.

On a purely numerical level, we note that this prescription coincides with the ``square root'' prescription for large distances $r$. In the latter situation, most of the contributing distances $|\mathbf{x}-\mathbf{y}|$ are smaller than $r$, hence reproducing the ``square root'' definition. On the contrary, for small separations $r$ in the correlation function, the scale of $\alpha_s$ will be set by that distance, thus modifying the final correlation function with respect to the ``square root'' prescription. Numerical confirmation of this expectation will be discussed in Section \ref{sec. conclusions}.

\section{Statistical analysis}
\label{sec. statistical analysis}

The final results, i.e. the two-point correlation functions of two Wilson lines, are evaluated using 100 random realizations of the color charge distributions. Standard statistical analysis provides standard deviations for the resulting data points.

For the definition of the saturation scale $Q_s$, we follow Ref.~\cite{Rummukainen:2003ns} and define it as the transverse momentum for which the rescaled gluon distribution reaches a maximum. The results are quoted in terms of the dimensionless quantity $L Q_s$. When plotted on a logarithmic scale, the gluon distribution near the maximum resembles a quadratic or a Gaussian function. Therefore, we use two fitting ansatzes in order to estimate the maximum:
\begin{eqnarray}
    f_{\textrm{quadratic}}(x) &=& a + b \ \big\{ \log(L k_T) - d \big\}^2 \,, \\
    f_{\textrm{gaussian}}(x) &=& a + b \exp \big\{ - c \big(\log(L k_T) - d \big)^2 \big\} \,,
\end{eqnarray}
with $a$, $b$, $c$ and $d$ being fit parameters. We associate the saturation scale with the parameter $d$. An example fit is shown in the uppermost panel of 
Fig.~\ref{fig. fitting}. The fitting range was chosen as $L k_T \in [33,245]$ and is shown in the figure as they gray band. In order to disentangle the discretization effects from the continuum dependence on $\log(L k_T)$, we increase the statistical uncertainties by a factor 3 in such a way that the data points form a continuous curve within their new uncertainties. In the example shown, 
both fitting functions describe the data well in the full range. The quality of the fits can be estimated by the value of $\chi^2/{\rm dof}$. In order to estimate the systematics of the fitting procedure, we vary the fitting range by shifting the left and right borders of the fitting interval and keeping the interval symmetric around the approximate maximum at $L k_T^{\textrm{max}} \approx 90$. The minimal range is $[55,148]$ and the maximal is $[12,665]$ and they contain respectively 50 and 250 data points. The middle panel of 
Fig.~\ref{fig. fitting} shows the dependence of the $\chi^2/{\rm dof}$ on the fitting range for both fitting ansatzes. Starting at $L k^{\textrm{initial}}_T \approx 20$, the $\chi^2/\textrm{dof}$'s of both kinds of fits are below 1 and approximately equal. The statistical uncertainty of the fitting parameters estimated using jackknife resampling from 100 fits is at a level of 1 \textperthousand \ and completely negligible compared to the systematic effects associated with the choice of the fitting range.

\begin{figure}
\begin{center}
\includegraphics[width=0.5\textwidth]{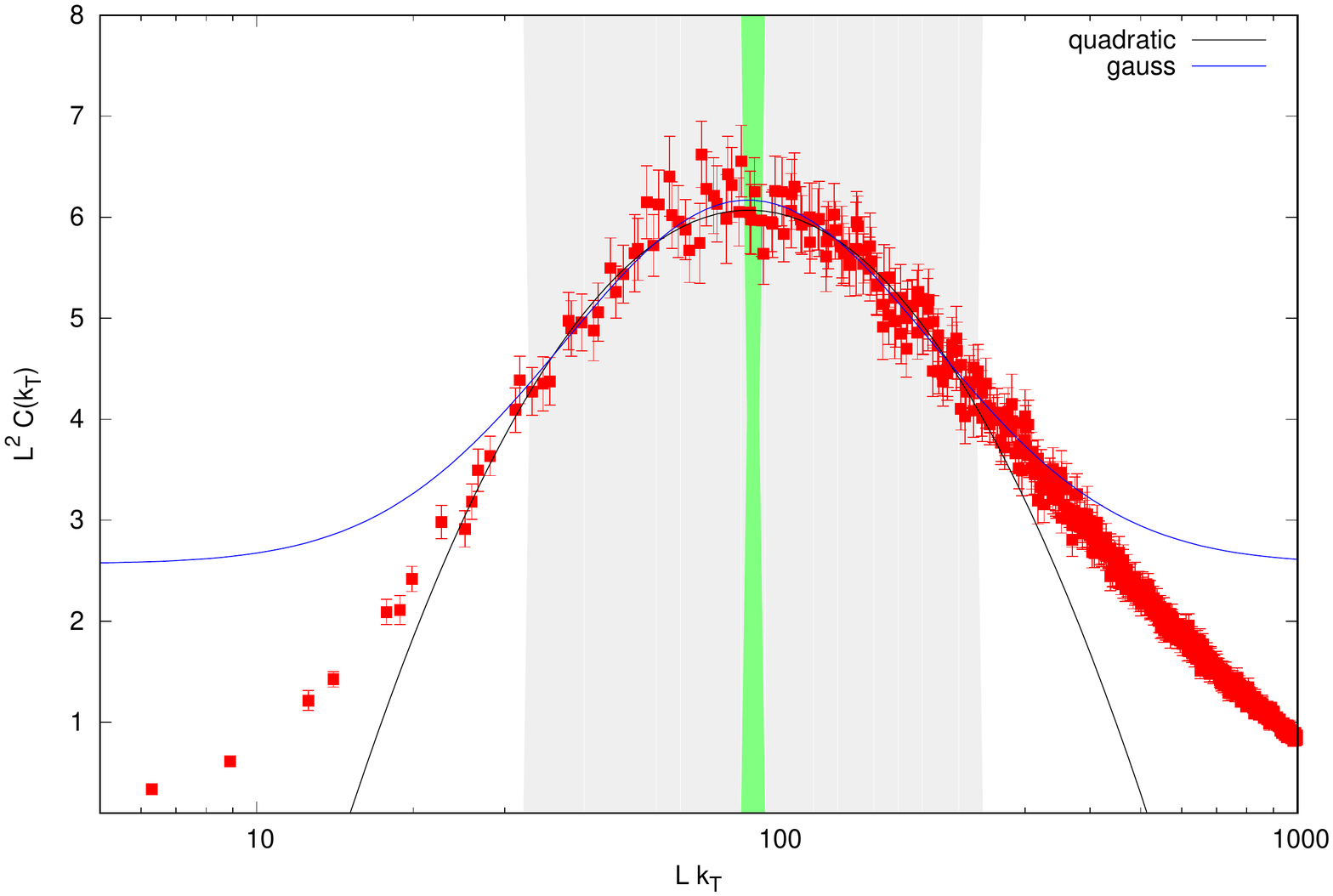}
\includegraphics[width=0.5\textwidth]{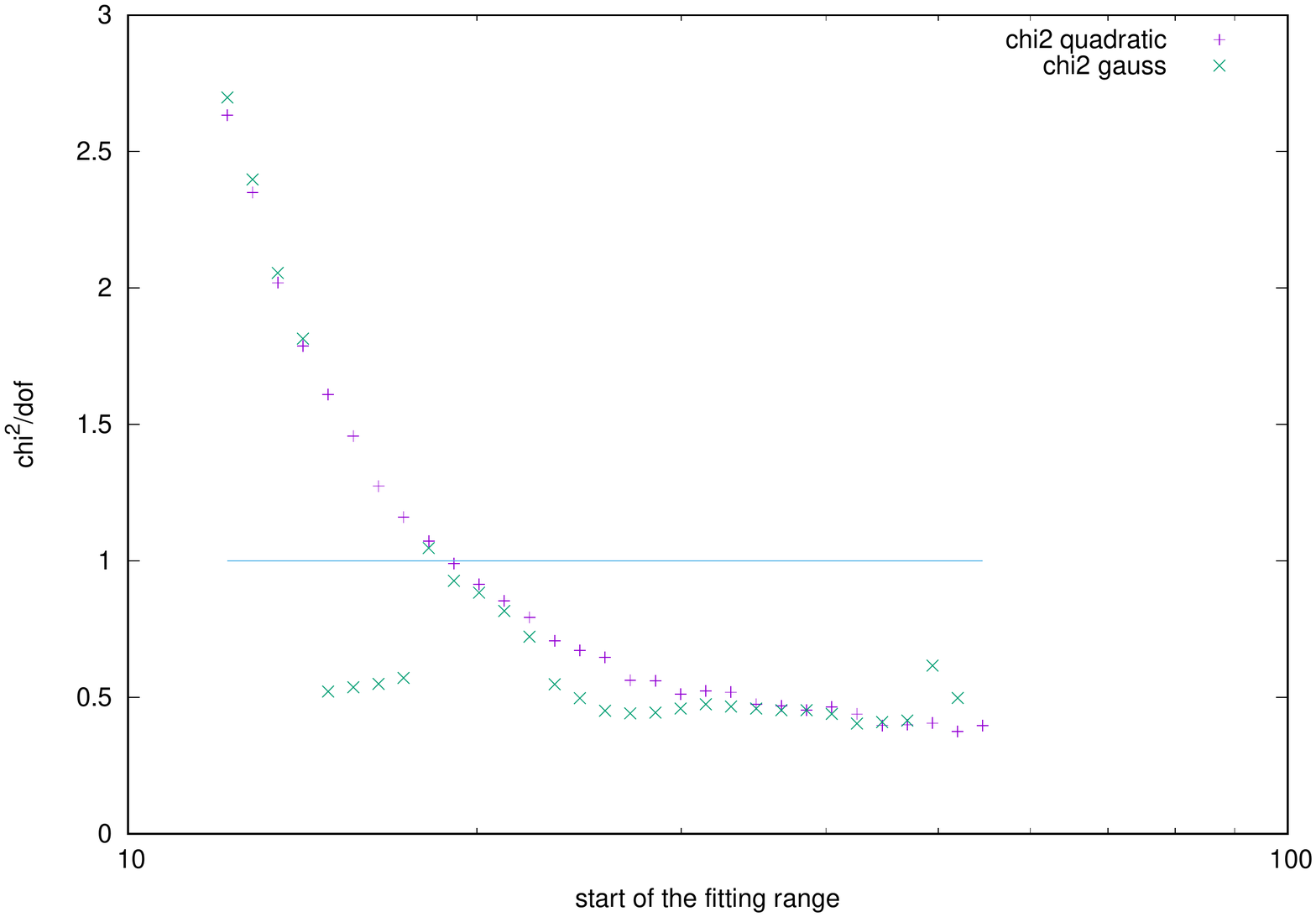}
\includegraphics[width=0.5\textwidth]{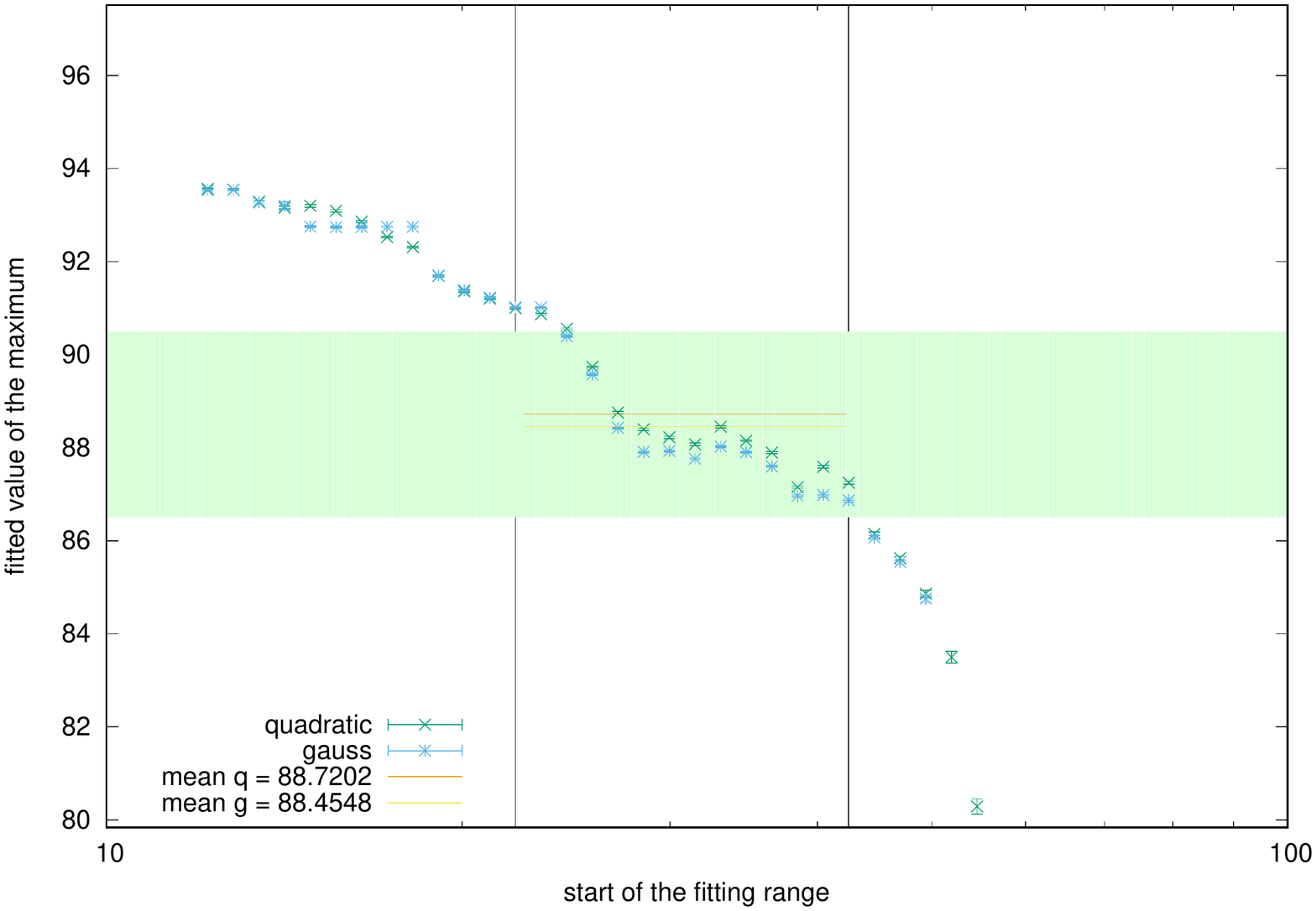}
\caption{Example of extracting the saturation scale $Q_s$ by fitting the maximum of the gluon distribution on a $512 \times 512$ lattice, at rapidity $s=0.04$, evolved in position space. The statistical errors were increased by a factor 3 in order to account for the discretization effects. The upper panel shows the comparison of the quadratic and Gaussian fits with the green band being the 1-$\sigma$ uncertainty of the saturation scale and the gray band showing the fitting range. The middle panel shows the value of $\chi^2/\textrm{dof}$ as a function of the fitting range. The bottom panel shows the dependence of the extracted maximum on the fitting range for both employed fitting ansatzes. \label{fig. fitting}}
\end{center}
\end{figure}

The final values of the saturation scale are shown in the bottom panel of 
Fig.~\ref{fig. fitting}. We exclude results for $L k^{\textrm{initial}}_T < 20$ as the fitting ansatzes do not correctly describe the tails of the correlator. We also exclude results for $L k^{\textrm{initial}}_T > 50$, as the fitting range contains too few data points to correctly resolve the maximum. For the remaining range of fitting intervals, we estimate the mean saturation scale obtained using the Gaussian ansatz as the final result, since the Gaussian curve describes the data for a wider range of $L k_T$ and we associate with it a systematic uncertainty equal to the spread of the remaining fit results coming from both ansatze. Hence, in the example shown, the final saturation scale reads $L Q_s = 88.5 \pm 2.0$. This corresponds to the green band in the top panel of Fig.~\ref{fig. fitting}.

\section{Systematics}
\label{sec. systematics}

This is the main section of our work, where 
we quantify and discuss the different systematic effects induced by the implementations outlined above. We start in Section \ref{sec. initial condition} with the discussion of the gluon distributions resulting from the initial condition (i.e.\ at zero rapidity) and their dependence on parameters such as volume, $N_y$ or the mass regulator in the Poisson equation. Next, in section \ref{sec. evolution}, we focus on the different ways the evolution in rapidity can be implemented. Eventually, in Section \ref{sec. running coupling systematics}, we discuss the systematic effects involved in the implementation of the running coupling.

\subsection{Initial condition}
\label{sec. initial condition}

\begin{figure}
\begin{center}
\includegraphics[width=0.5\textwidth]{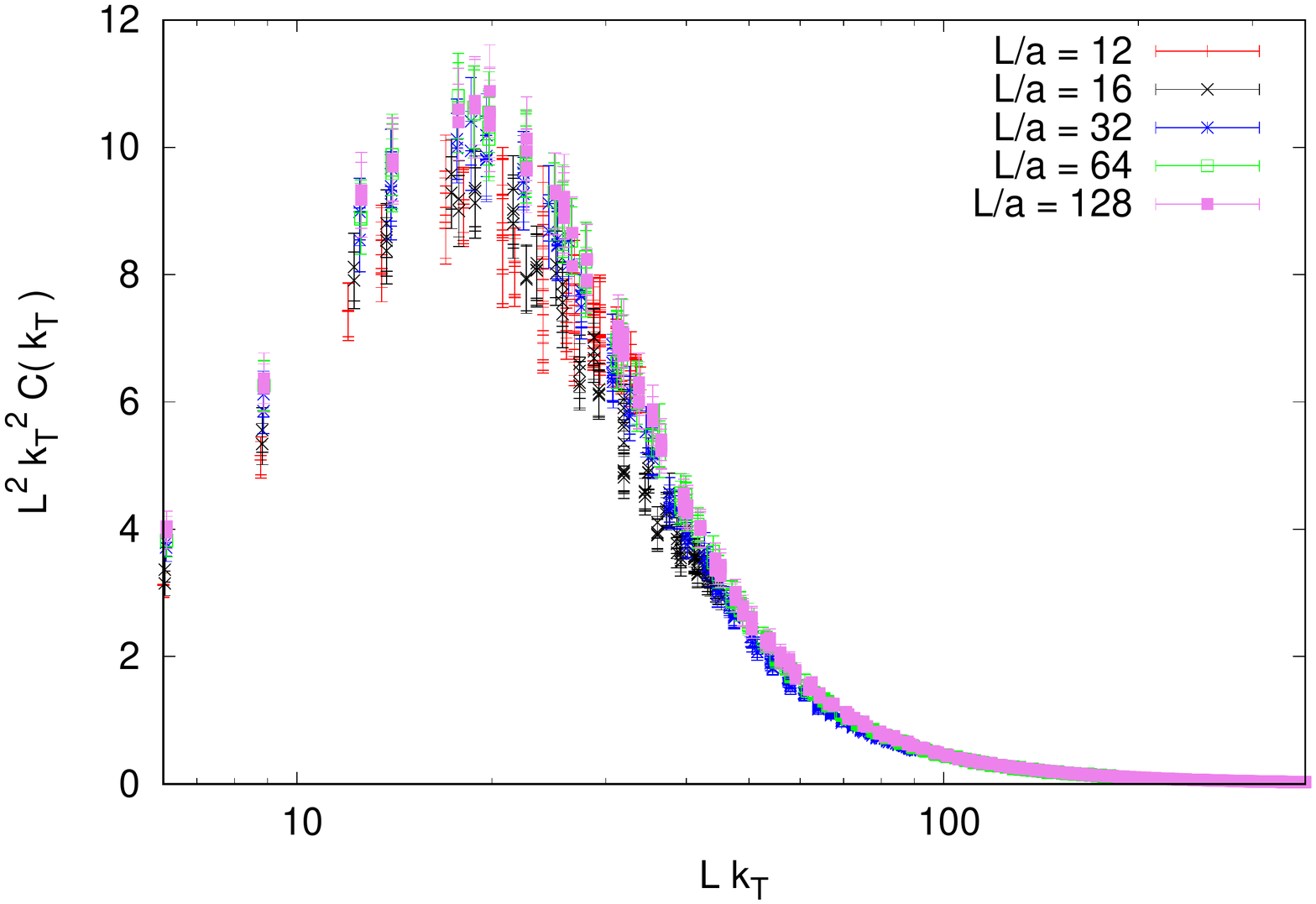}
\includegraphics[width=0.5\textwidth]{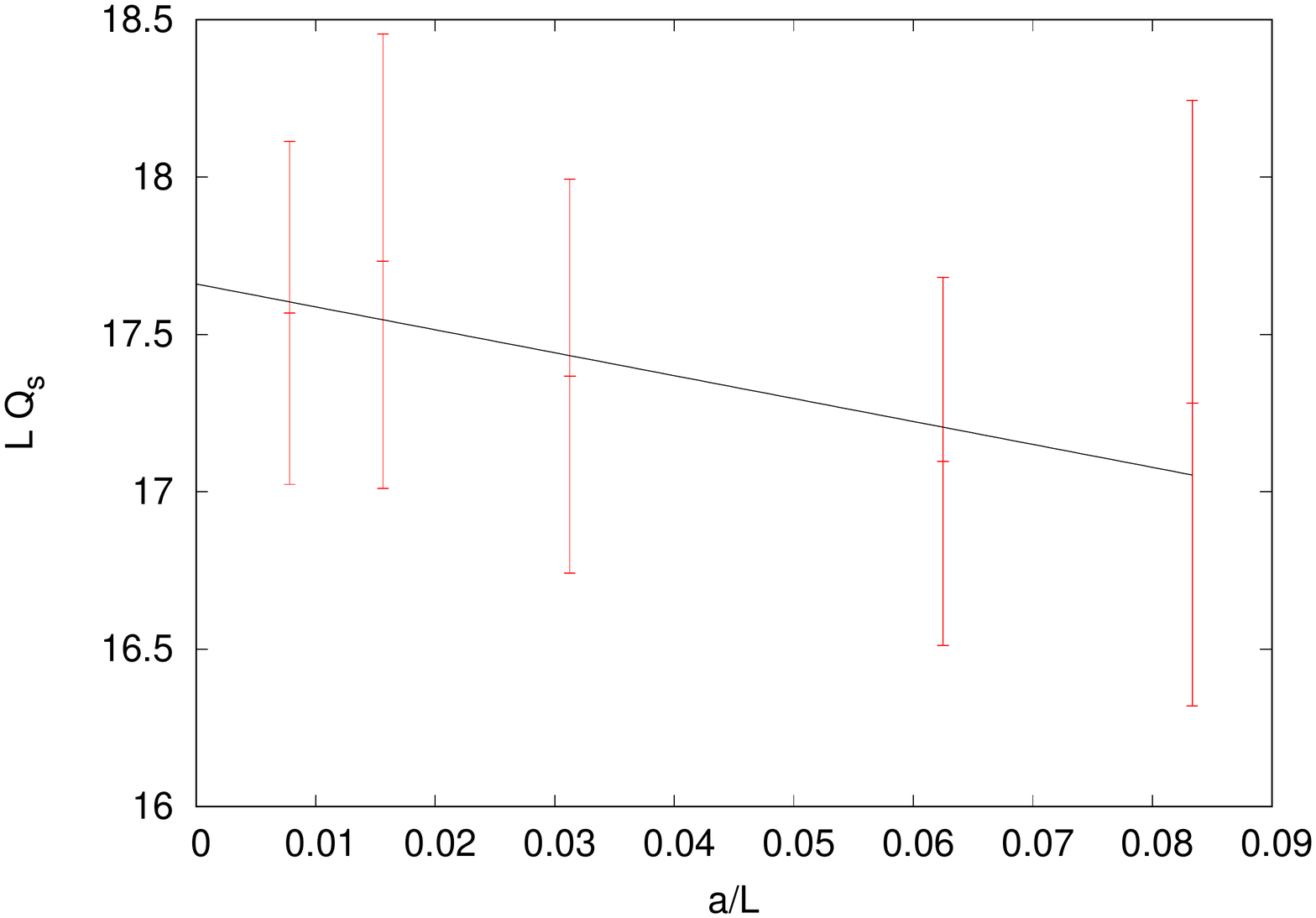}
\caption{Top: Volume dependence of the momentum-space correlator from the initial condition. Bottom: Dependence of the saturation scale on the lattice extent. Black line is the attempted infinite volume extrapolation with $f(1/L) = a + b/L$. Clearly all the volumes are compatible with the extrapolated value, hence no finite volume effects are associated with the initial condition at these values of the MV model. \label{fig. initial condition volume}}
\end{center}
\end{figure}
\begin{figure}
\begin{center}
\includegraphics[width=0.5\textwidth]{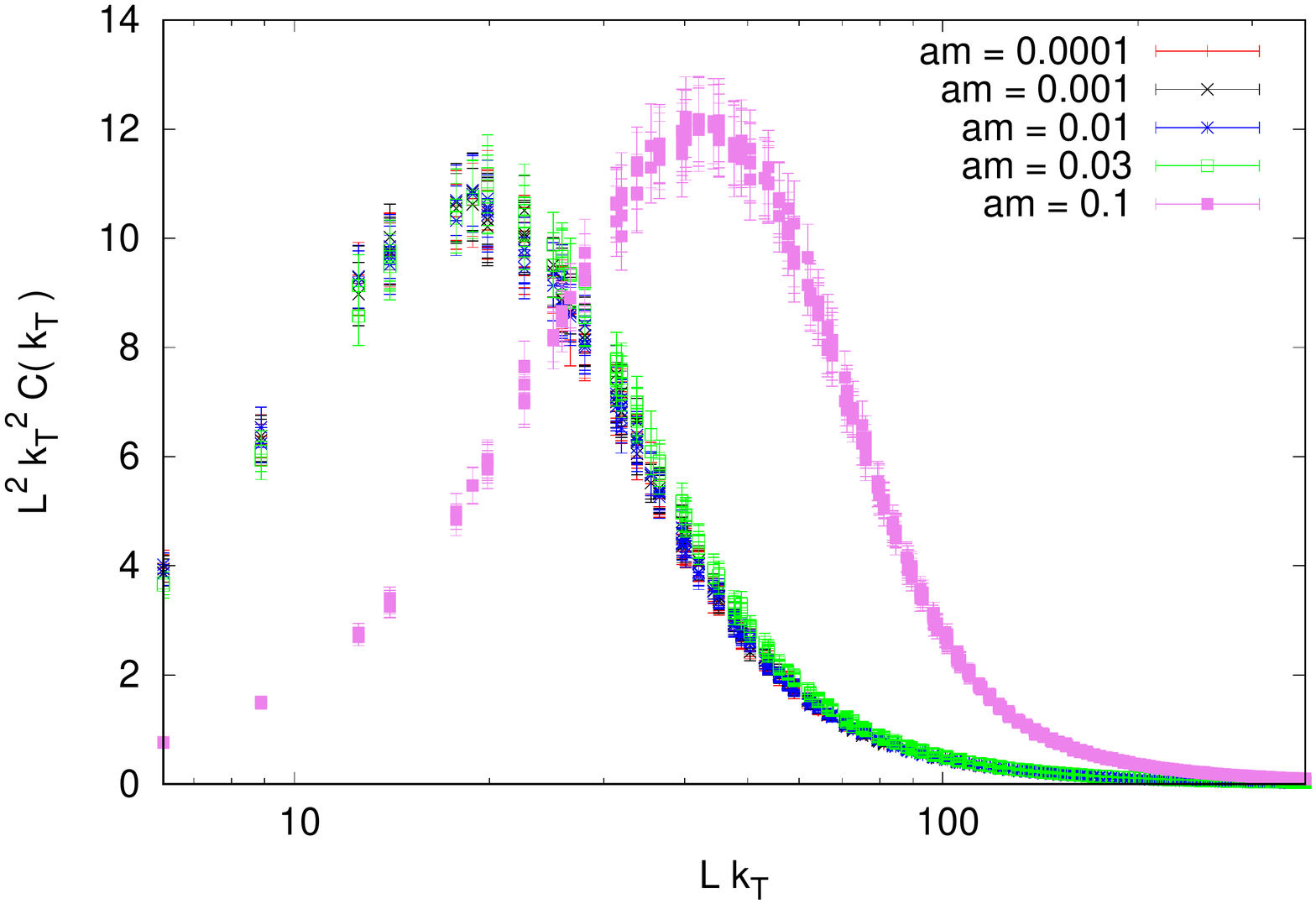}
\includegraphics[width=0.5\textwidth]{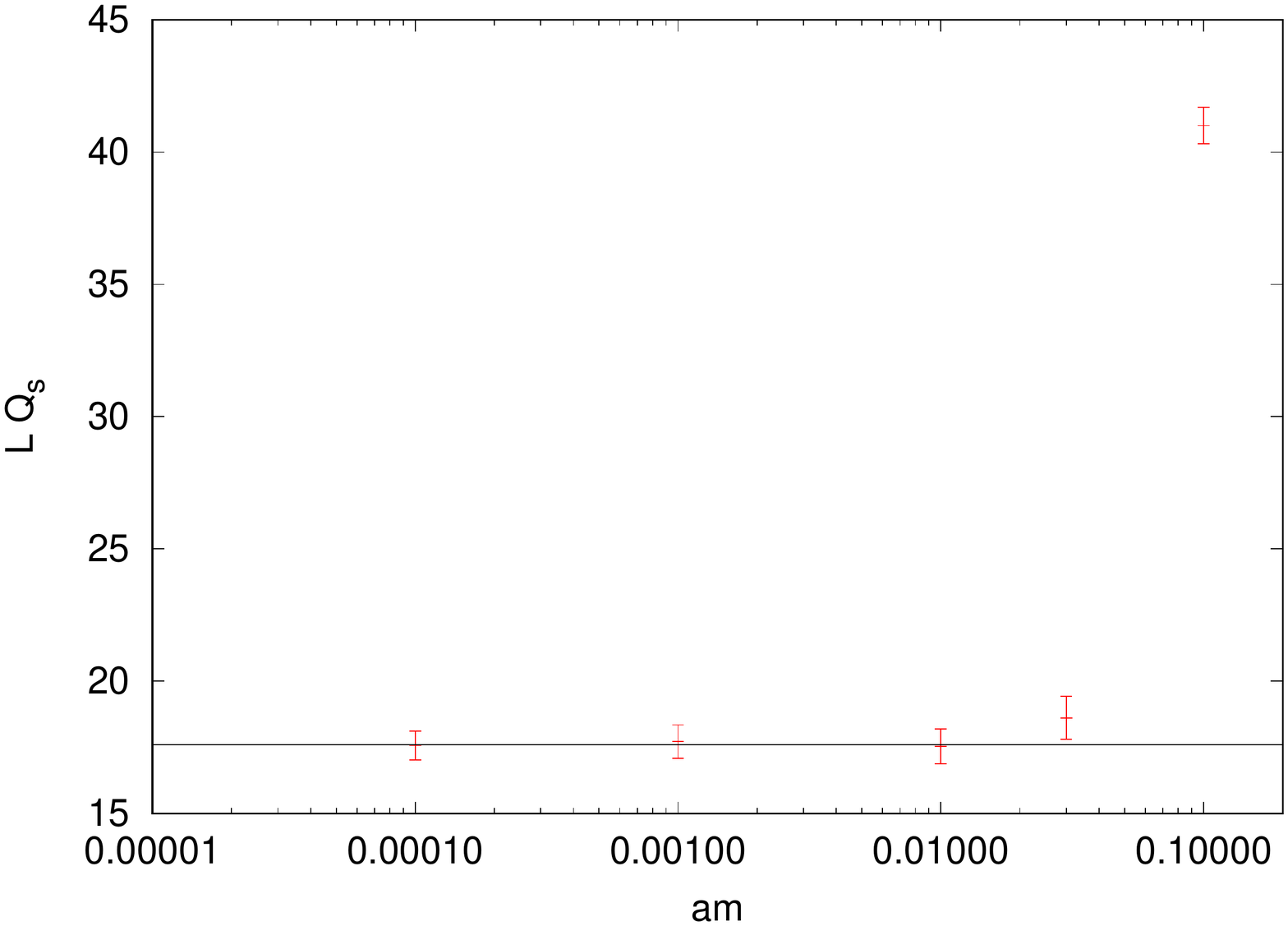}
\caption{Top: Dependence of the $s=0$ momentum-space correlator on the $m$ regulator in Eq.~\eqref{eq. poisson}. Bottom: Dependence of the saturation scale on the mass regulator. A vanishing mass extrapolation is attempted with the constant fit ansatz $f(am) = b$, where $b$ is a fit parameter. Data points with $am \sim 0.03$ are still compatible with the extrapolated value. \label{fig. initial condition mass}} 
\end{center}
\end{figure}
\begin{figure}
\begin{center}
\includegraphics[width=0.5\textwidth]{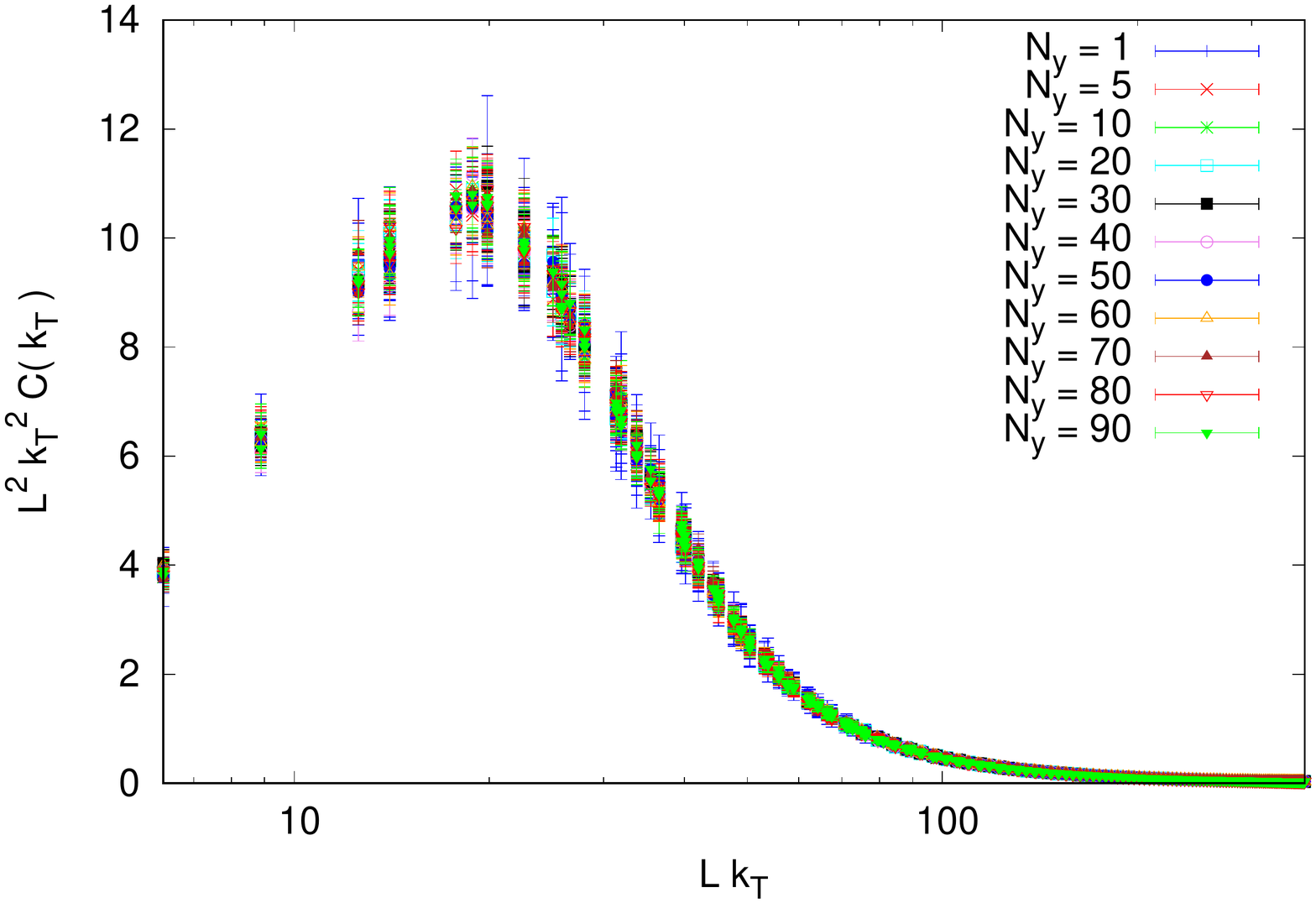}
\includegraphics[width=0.5\textwidth]{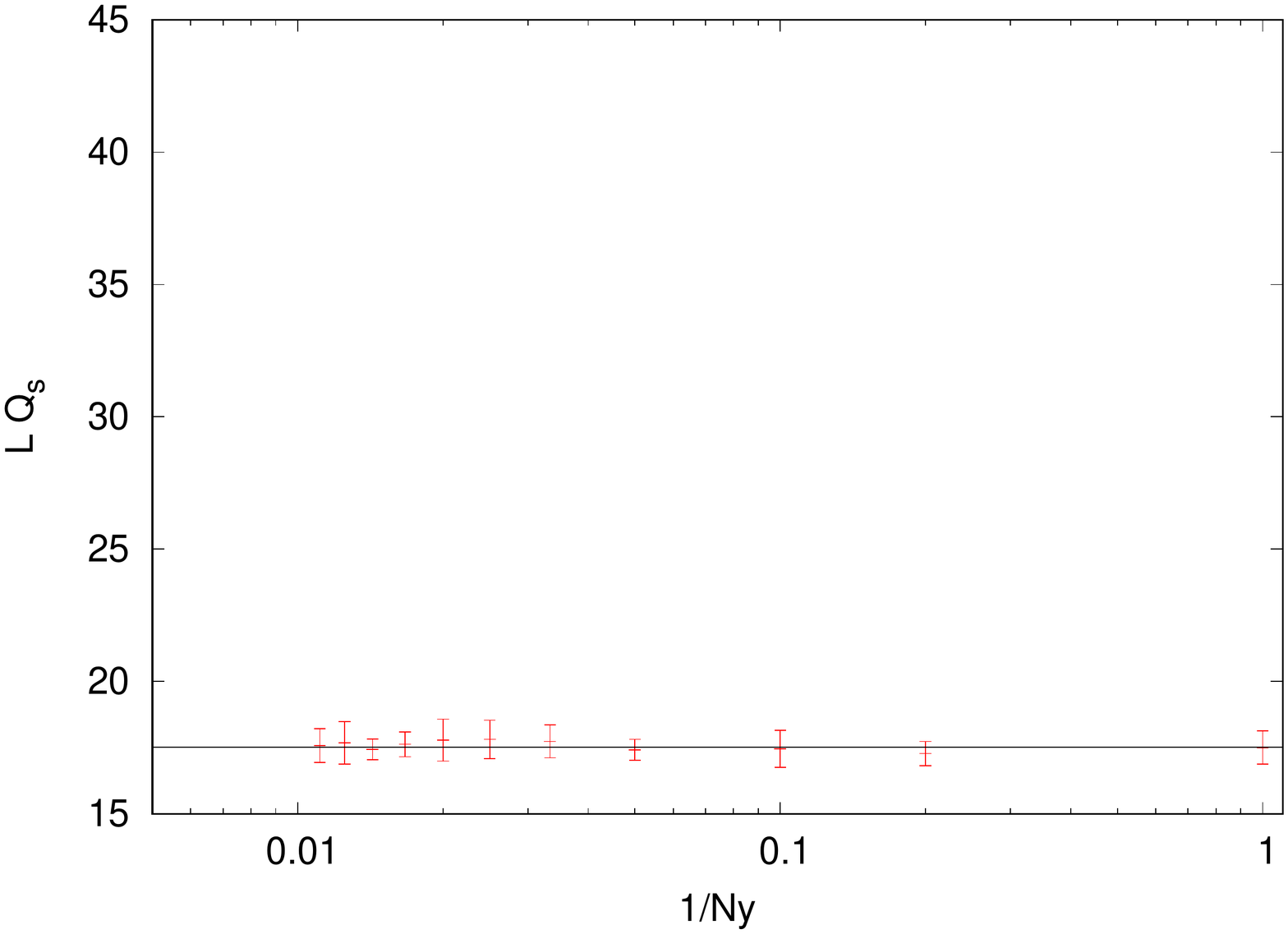}
\caption{Top: Dependence of the $s=0$ momentum-space correlator on the $N_y$ parameter. Bottom: Dependence of the saturation scale on $N_y$. An infinite $N_y$ extrapolation is attempted with the constant fit ansatz $f(1/N_y) = b$, where $b$ is a fit parameter. Even the data point $N_y=1$  is compatible with the extrapolated value. \label{fig. initial condition Ny}} 
\end{center}
\end{figure}

In this section, we discuss the dependence of the two-point correlation function on the parameters of the numerical setup. The parameters of the MV model have to be scaled appropriately when the lattice size is varied. We keep the dimensionless combination $g^2\mu L$ constant as we scan the different lattice extents from $L/a=12$ up to $L/a=128$. We use the value $g^2\mu L=30.72$ from Refs.~\cite{Lappi:2012vw,Marquet:2016cgx}. 
As far as the lattice volume is concerned, we do not see any statistically significant deviations down to lattice sizes of $L/a=12$, see Fig.~\ref{fig. initial condition volume}. The dependence on the mass regulator in the Poisson equation, Eq.~\eqref{eq. poisson}, is shown in Fig.~\ref{fig. initial condition mass} together with the dependence of the resulting saturation scale. We notice that within the systematic and statistical errors, all setups provide consistent saturation scales, except the single case of the largest mass regulator, $am=0.1$. 
Finally, the dependence on $N_y$ is shown in Fig.~\ref{fig. initial condition Ny}. We check that when a statistical ensemble of 256 realisations is used, all values of $N_y$ saturate the product in Eq.~\eqref{eq. wilson line}. Since the generation of the initial distribution is a negligible cost of the computation compared to the evolution in rapidity, we conservatively use the value of $N_y=50$ for all subsequent calculations.

\subsection{Evolution at fixed coupling}
\label{sec. evolution}

\begin{figure}
\begin{center}
\includegraphics[width=0.5\textwidth]{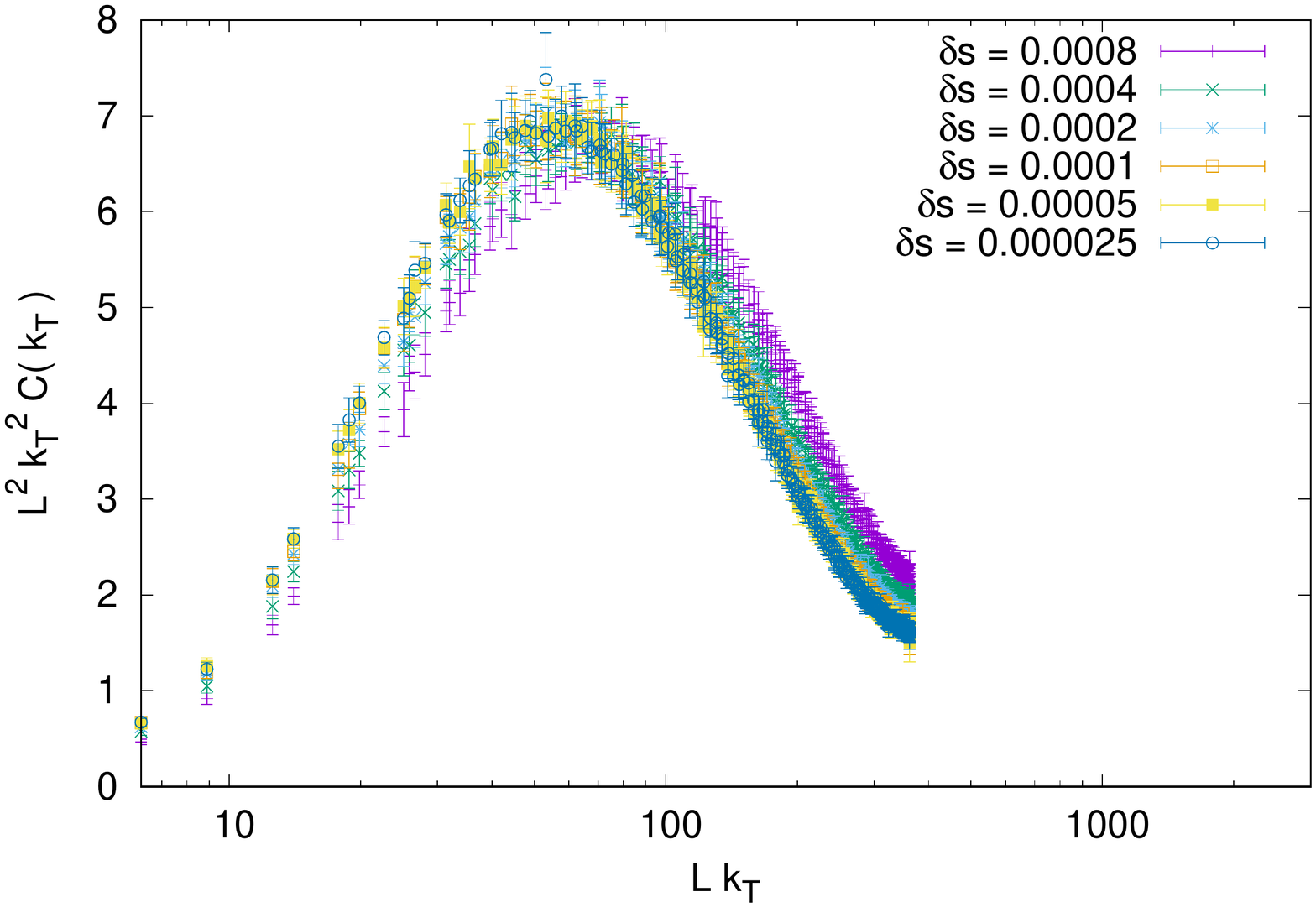}
\includegraphics[width=0.5\textwidth]{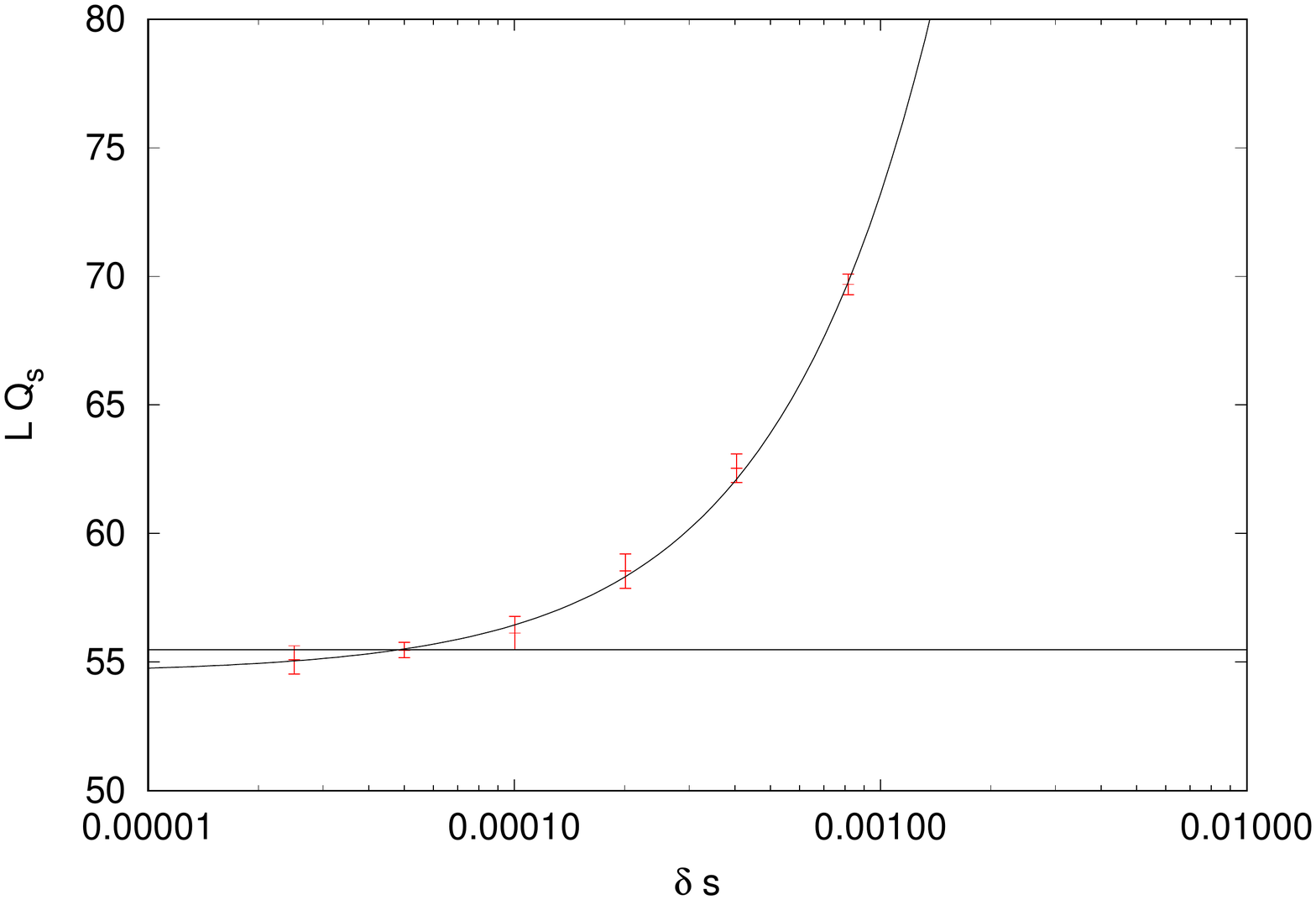}
\caption{Top: Dependence of the evolved two-point correlation function at $s=0.04$ on the Langevin step $\delta s$ for the $L/a=128$ lattice, using the momentum space implementation of the evolution equation. Bottom: 
Dependence of the saturation scale determined from the data sets show in the top panel on the Langevin time step $\delta s$. Constant and linear extrapolations to the vanishing step size are shown (note the logarithmic scale on the x-axis). For step sizes smaller than $\delta s = 0.0001$, data at finite values of the time step are compatible with the extrapolated values.
\label{fig. evolution position timestep}}
\end{center}
\end{figure}
\begin{figure}
\begin{center}
\includegraphics[width=0.5\textwidth]{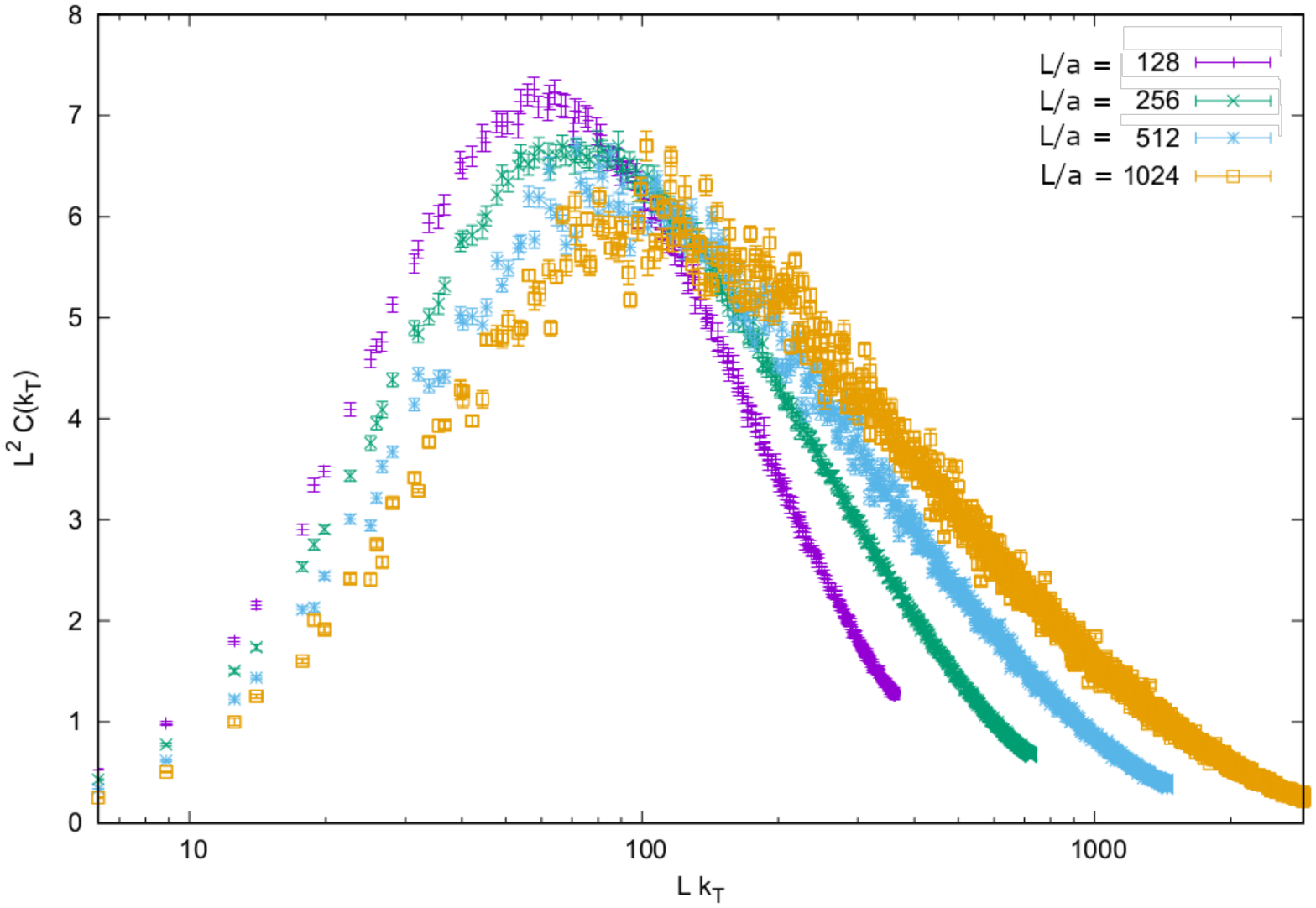}
\includegraphics[width=0.5\textwidth]{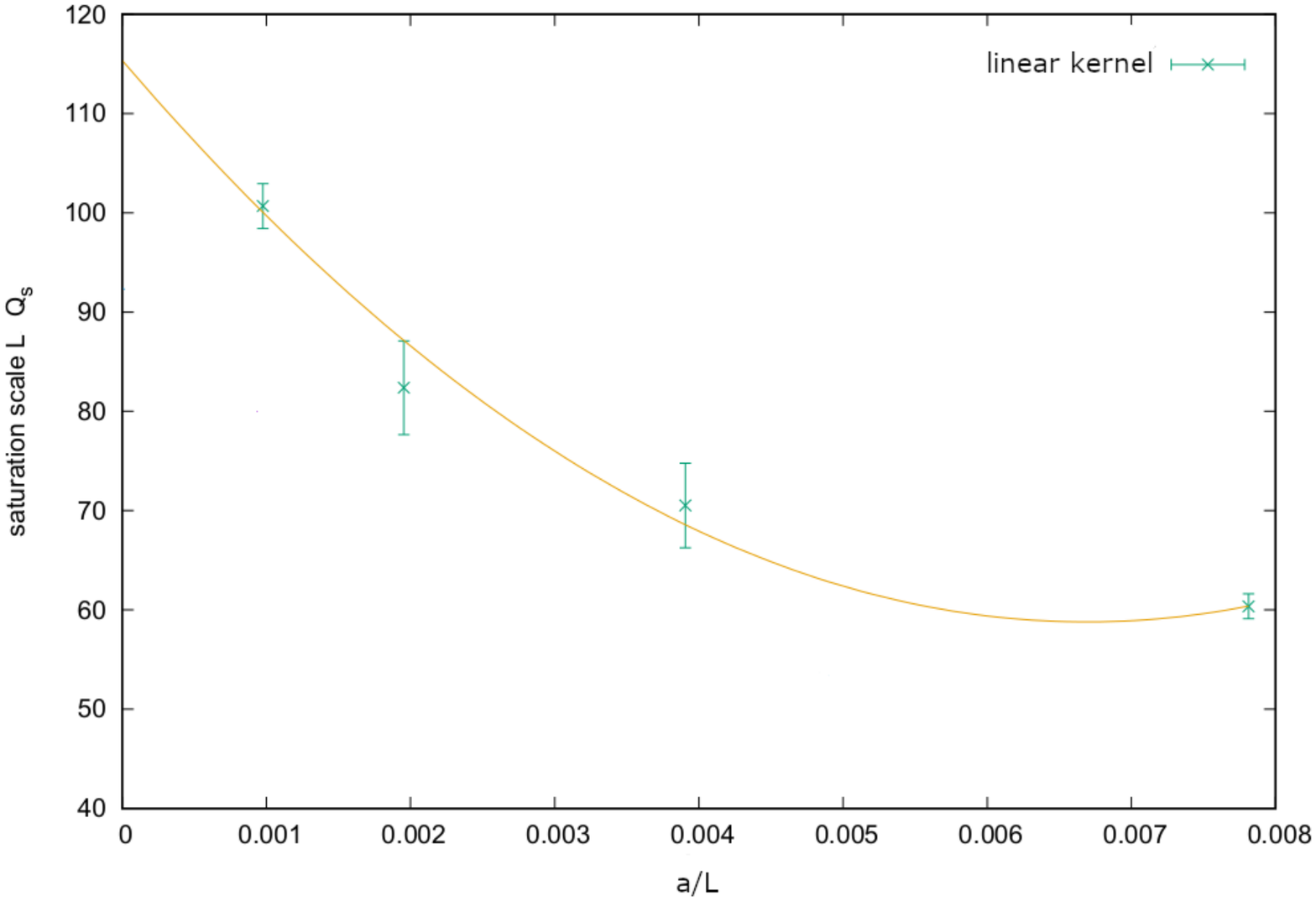}
\caption{Comparison of the evolved two-point correlation function as a function of the volume. Evolution to $s=0.04$ with time step $\delta s = 0.0001$. Infinite volume extrapolation is attempted demonstrating the saturation scale divergence in the volume parameter.  The fit ansatz is $f(a/L) = b + c \frac{L}{a}$ with $b$ and $c$ fit parameters. Similar behaviour was found for the sine kernel. \label{fig. position space evolution volume dependence}}
\end{center}
\end{figure}

In this section, we study systematic effects which arise from the different implementations of the Langevin equation. As we will see, the situation at fixed coupling constant is not satisfactory, as the continuum limit is not well-defined. Therefore, we do not present here all the possible implementation combinations and only concentrate on some representative features of the fixed-coupling setup. The phenomenologically relevant situation with the running coupling constant is discussed in the next section.

We study the dependence of the gluon distributions on various parameters, in particular the size of the time step and the lattice extent and the different discretizations of the JIMWLK kernel. In the top panel of 
Fig.~\ref{fig. evolution position timestep}, we show the dependence of the distributions on the size $\delta s$ of the Langevin step in the evolution equation. We show data for rapidity $s=0.04$ together with the initial condition to set the scale. In the fixed coupling scenario, we remind that one can relate $s$ to the rapidity $y$ by \cite{Marquet:2016cgx}
\begin{equation}
    s = \frac{\alpha_s}{\pi^2} y.    
\end{equation}
Therefore, taking e.g.\ $\alpha_s=0.16$, $s=0.04$ corresponds to $y \approx 2.5$.
For each value of the time step, we estimate the saturation scale and its systematic and statistical error and plot it as a function of the time step in the bottom panel of Fig.~\ref{fig. evolution position timestep}. Two extrapolations to the vanishing value of the Langevin time step are shown in the figure: 
i) a constant one, fitted to the last three data points, which are compatible within their systematic uncertainties,
ii) a linear one in the time step, fitted up to $\delta s = 0.0008$. Both extrapolations yield values which are compatible within their uncertainties. Hence, we notice convergence. For time steps smaller than $\delta s = 0.0001$, there is no systematic difference in the saturation scale and the results at non-zero step size are compatible within uncertainties with the extrapolated results. 

Another important parameter is the lattice extent $L/a$. Its impact on the evolution is shown in 
Fig.~\ref{fig. position space evolution volume dependence}. We present data for different volumes for the evolution with time step of $\delta s = 0.0001$ and for final rapidity of $s=0.04$. We see that the results significantly depend on the lattice volume for volumes in the range from $L/a=128$ up to $L/a=1024$. When analyzing the resulting saturation scales, a systematic trend is clearly visible with the saturation scale diverging as the infinite volume limit is taken. We find the same behaviour in both position and momentum space formulations, where, for the latter, we were able to reach volumes as large as $L/a=16384$. This observation is also irrespective of the fact whether the 'linear' or 'sin' kernel is used. This result is in agreement with the previous findings discussed in Ref.~\cite{Rummukainen:2003ns}. The difficulty in taking the continuum limit makes the evolution at fixed coupling not a viable numerical approach, hence we do not discuss it further here and we turn our attention to the case of evolution with running coupling.

Before doing so, we provide a consistency check of the implementations of the coupling, shown in Fig.~\ref{fig. evolution fixed coupling}. Namely, we compare evolved gluon distributions with a fixed coupling and with a running coupling, having switched off the running. Appropriate rescaling of noise vectors guarantees that the evolved distributions agree.

\begin{figure}
\begin{center}
\includegraphics[width=0.5\textwidth]{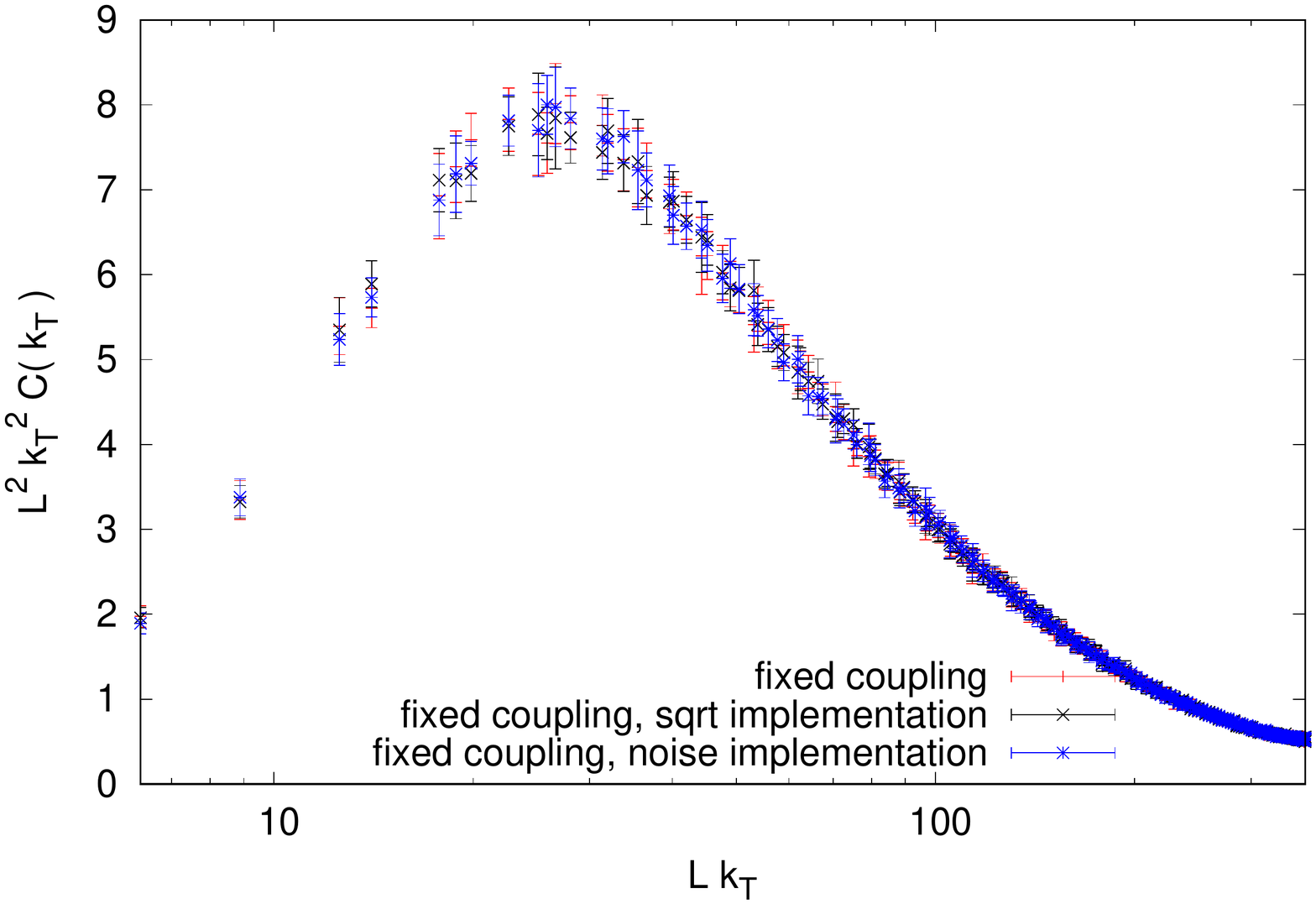}
\caption{Consistency check of the implementation of coupling prescriptions: the fixed coupling constant evolution at rapidity $s=0.08$ is compared with the results obtained with running coupling constant with the running switched off. Agreement of the curves shows that although different implementations are used, all normalization factors are correctly taken into account.\label{fig. evolution fixed coupling}} 
\end{center}
\end{figure}

\subsection{Evolution at running coupling}
\label{sec. running coupling systematics}

Now, we turn on the effects arising from the running of the coupling. As discussed in Section~\ref{sec. running coupling}, there are several possible ways to implement the latter and here, we compare the systematics arising in the ``square root'' and ``noise'' prescriptions. 

Let us start by noticing that in the setup where the running coupling constant is included, the numerical evolution proceeds much slower. 
This results from the fact that now, the Langevin step contains a scale-dependent $\alpha_s$ factor. 
For example, the value of $s = 0.04$ that we used to illustrate the systematics of the fixed-coupling setting, corresponds in the case of a running coupling to $y \approx 0.4$ (taking $\alpha_s$ evaluated at the saturation scale) instead of $y\approx2.5$. Thus, in order to reach the same physical rapidity, the simulations with a running coupling must be considerably longer.
We will comment more on the speed of the evolution with different prescriptions for $\alpha_s$ in the next section. 

\begin{figure}
\begin{center}
\includegraphics[width=0.5\textwidth]{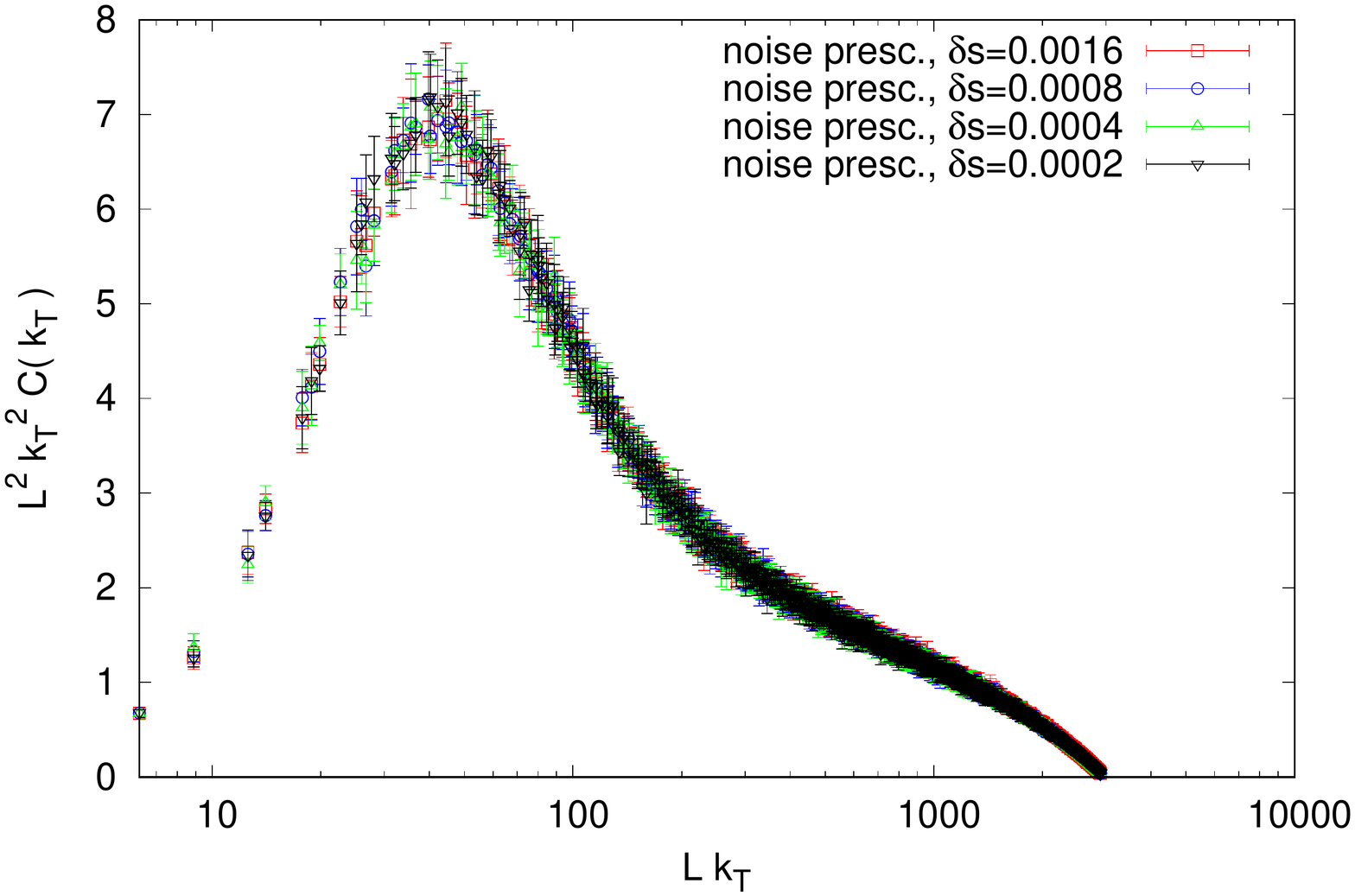}
\vspace{-0.5cm}
\includegraphics[width=0.5\textwidth]{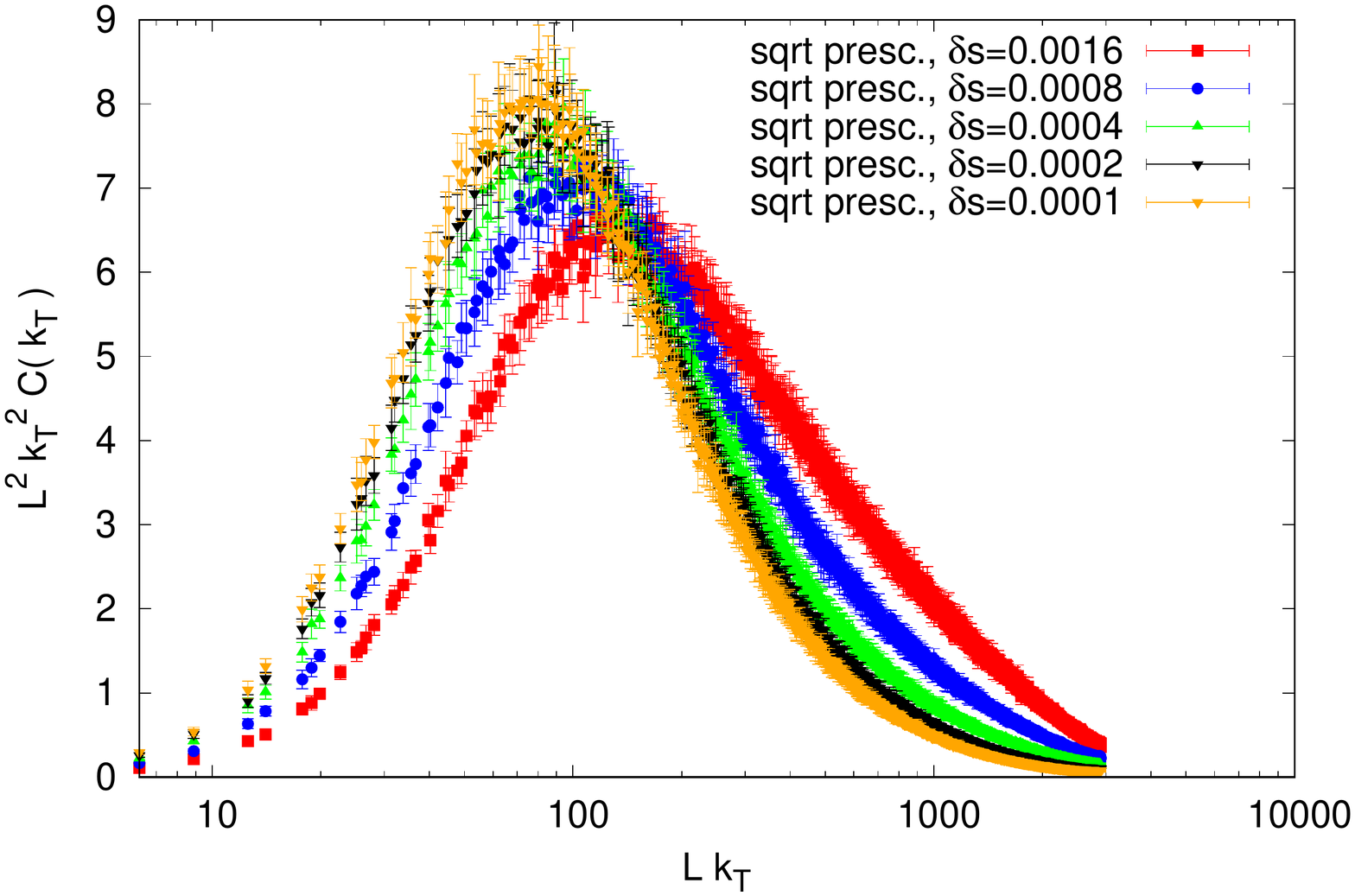}
\vspace{-0.5cm}
\includegraphics[width=0.5\textwidth]{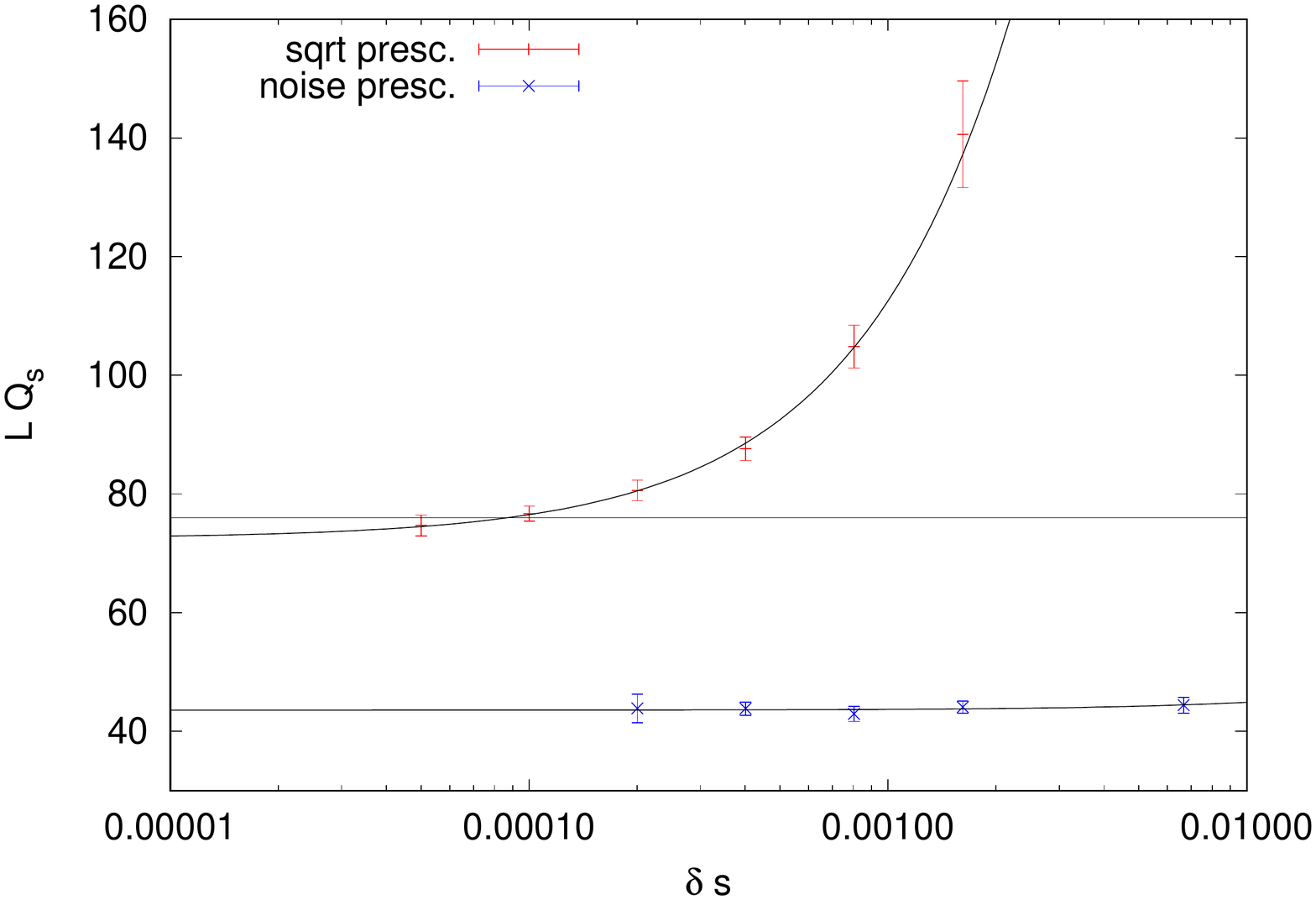}
\vspace{-0.5cm}
\caption{Comparison of gluon densities obtained with the ``noise'' (top) and ``square root'' (middle) prescriptions for $\alpha_s$ as a function of the Langevin step size at rapidity $s=0.16$. 
Bottom: vanishing step size extrapolation of the saturation scale $Q_s$ determined from the maxima of the distributions. We used a simple fit ansatz of the form $LQ_s(\delta s) = b + c \delta s$, with $b$ and $c$ being fit parameters. One sees a dramatic difference in the convergence rate: the ``noise'' prescription is roughly insensitive to the step size in the investigated range, whereas the ``square root'' prescription exhibits a linear slope similar to the one found in the fixed coupling setting. A constant fit to the two finest step sizes ($\delta s\leq0.0001$) for the ``square root'' data was also performed, showing that the continuum limit has been attained. The data set corresponding to the finest step size, $\delta s=0.00005$ is not shown on the middle panel, as it overlaps almost ideally with $\delta s=0.0001$. \label{fig. running coupling step size}}
\end{center}
\end{figure}

We start with the discussion of the dependence on the size of the Langevin step in the numerical integration of the Langevin equation. 
We note that there is a significant difference between the two running coupling prescriptions, see Fig.~\ref{fig. running coupling step size}.
The Langevin step dependence of the ``square root'' prescription is similar to the one in the fixed coupling setting (cf.\ Fig.~\ref{fig. evolution position timestep}), with a similar slope, e.g.\ the result at $\delta s=0.0004$ is around 10-15\% above the linearly extrapolated one and one needs to go to $\delta s\leq 0.0001$ to ensure agreement with the latter.
In turn, the ``noise'' prescription allows for a step size at least one order of magnitude larger than the ``square root'' prescription and even the result at our largest employed step size is consistent with the extrapolated value. The origin of this effect was not identified, as both implementations use the same Wilson line update algorithm, and will be investigated in the future.
However, the practical conclusion is that much larger rapidities can be attained with the same computational cost with the ``noise'' prescription.

As a second systematic effect, we consider the volume dependence of the gluon densities obtained with different $\alpha_s$ prescriptions at a fixed rapidity of $s=0.16$. The results are shown in Fig.~\ref{fig. running coupling volume}. The upper two panels show the comparison of distribution shapes for both the ``square root'' and ``noise'' prescriptions.
We note that, contrary to the fixed coupling setting (see Fig.~\ref{fig. position space evolution volume dependence}), the peak of the distributions, i.e.\ the saturation scale, does not depend on the volume, provided the latter is large enough.
Differences between the shown volumes are visible only in the tails of the distributions and can be attributed to discretization effects -- in a fixed physical volume, smaller lattice volumes correspond to larger lattice spacings.
Even though saturation scales implied for the chosen parameter values are different for the two prescriptions, the systematics of the volume is the same, i.e.\ $L/a\geq256$ ensures that the saturation scale becomes independent of $L/a$ within uncertainties.
This is illustrated in the bottom panel of Fig.~\ref{fig. running coupling volume}, where we show the volume dependence of the saturation scale $Q_s$ determined from the maxima of the distributions.
In order to contain both sets of data on the same extrapolation plot, the ``square root'' data used for the extrapolation are taken at a twice smaller rapidity, $s=0.08$. 
The data can be extrapolated to the infinite volume limit using a constant fitting ansatz when the smallest two volumes used, namely $L/a=64$ and $L/a=128$ are removed from the fit. At this value of rapidity, they are affected by finite volume effects and hence are not reliable.

\begin{figure}
\begin{center}
\includegraphics[width=0.5\textwidth]{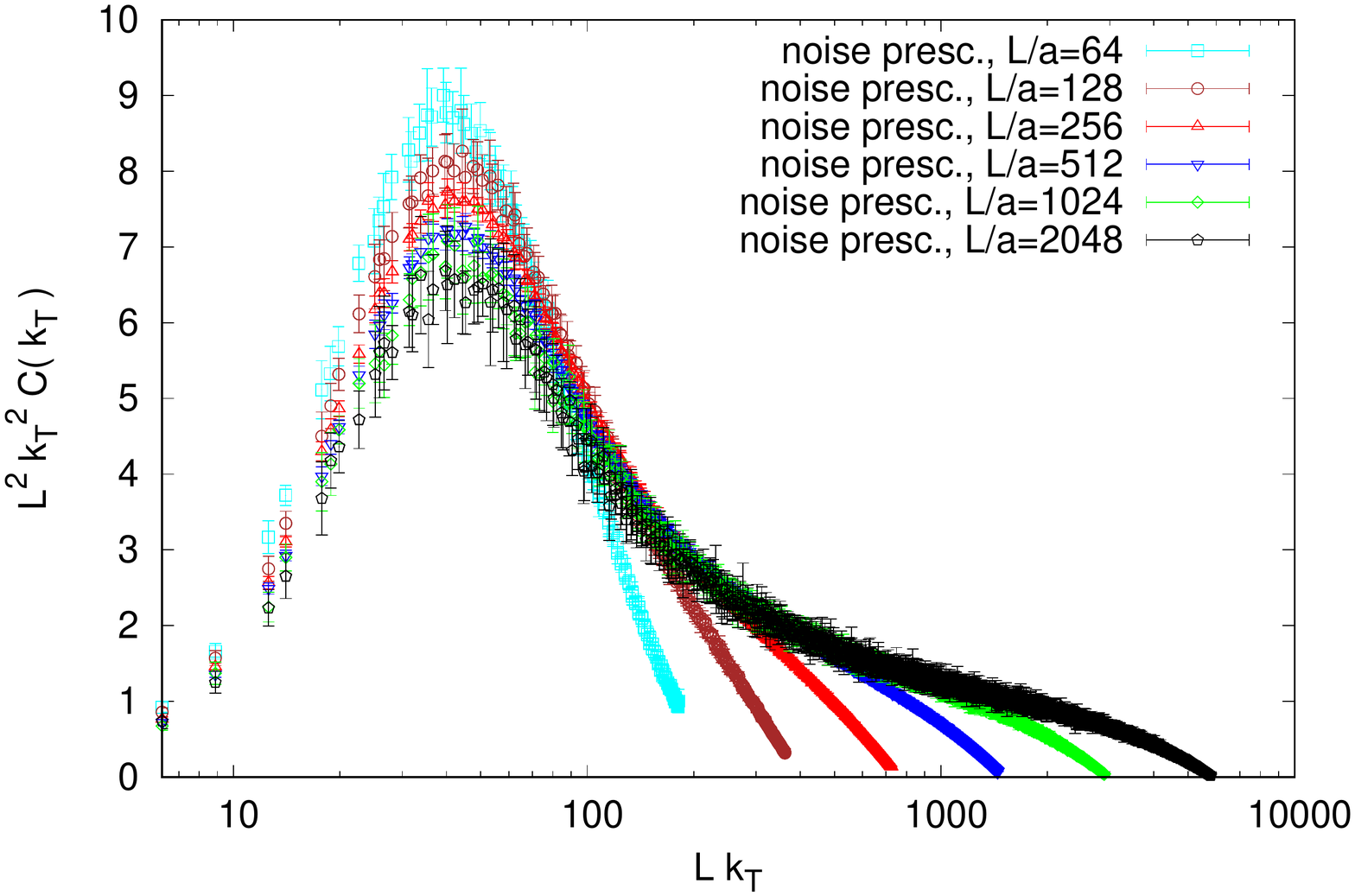}
\includegraphics[width=0.5\textwidth]{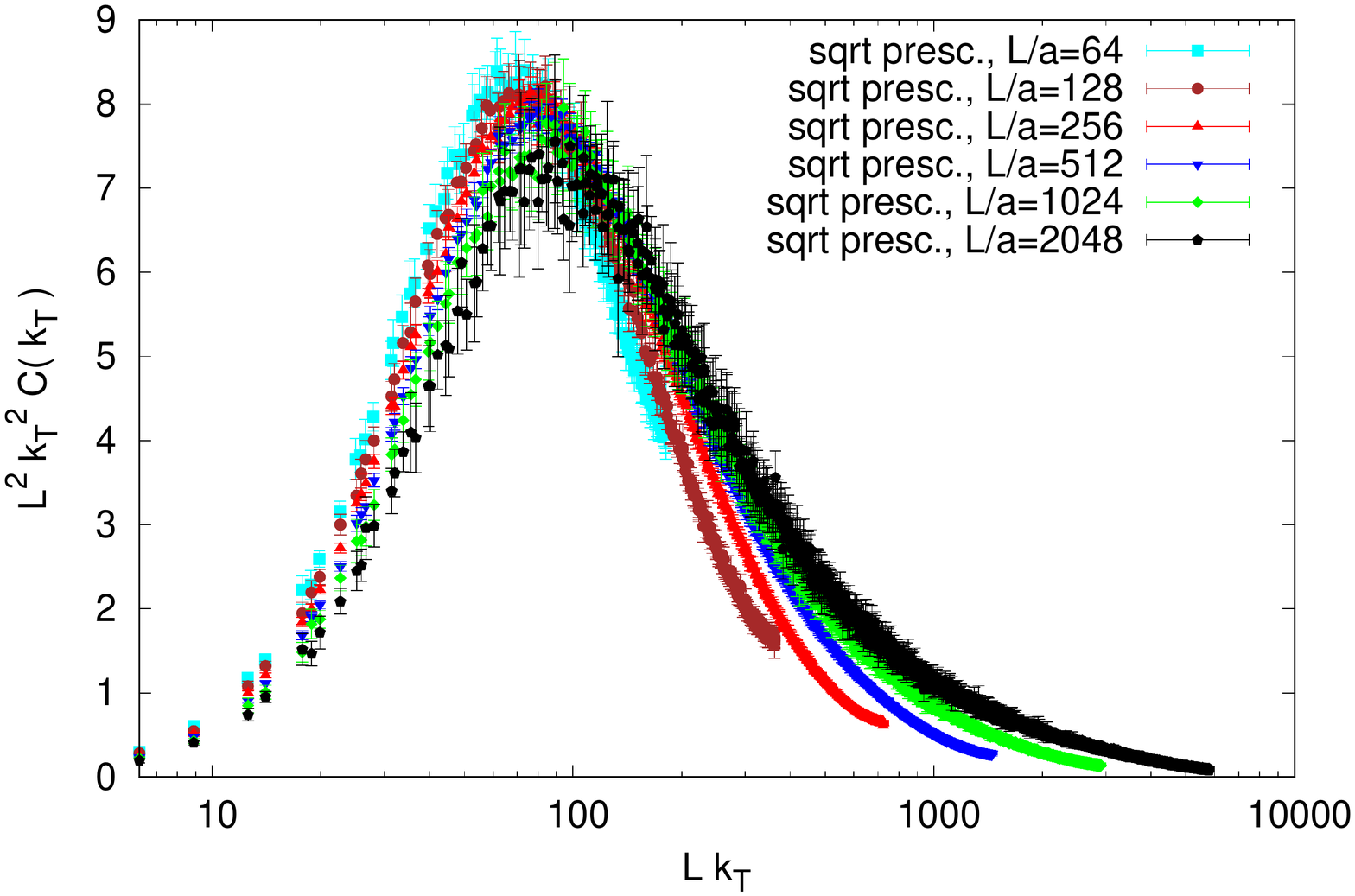}
\includegraphics[width=0.5\textwidth]{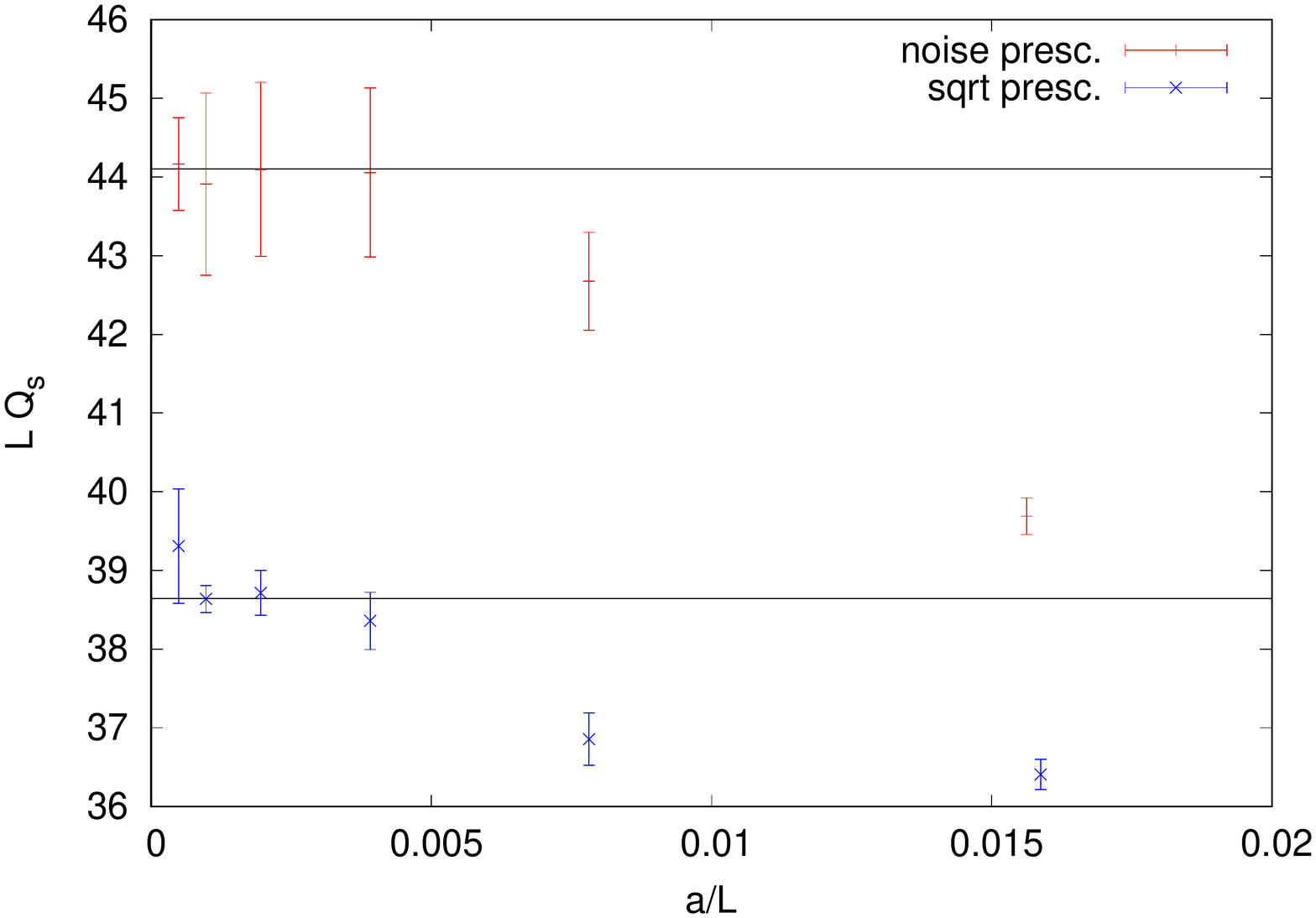}
\caption{Comparison of gluon densities obtained with the ``noise'' (top) and ``square root'' (middle) prescriptions for $\alpha_s$ as a function of the volume at rapidity $s=0.16$. 
Bottom: infinite volume extrapolation of the saturation scale $Q_s$ determined from the maxima of the distributions. We use a constant fit ansatz of the form $LQ_s(a/L) = b$ with $b$ being fit parameters, where the fitted data include the 4 largest volumes. In order to contain both sets of data on the same plot, the ``square root'' data are shown for a twice smaller rapidity, $s=0.08$. For volumes $L/a\geq256$, a constant extrapolation to the infinite volume limit describes the data well.  \label{fig. running coupling volume}}
\end{center}
\end{figure}

As a next step, Fig.~\ref{fig. running coupling kernel} shows the comparison of gluon distributions obtained using different JIMWLK kernel discretizations. As discussed above, we proposed to discretize the kernel in two ways, both in position space and in momentum space. The figure only shows results obtained in momentum space. The outcomes in position space follow a similar pattern. Comparing the gluon distributions obtained with the kernel using the sine function (``sin kernel'') and the naive discretization (``linear kernel''), we see that the latter induces large distortions for momenta $L k_T$ near the maximal value. In the case of the ``noise'' prescription, these distortions are limited to large $L k_T$ values, as opposed to the ``square root'' prescription, where the modifications also affect the position of the maximum of the gluon distribution and hence the saturation scale itself. As the results obtained with the linear kernel exhibit an unphysical growth of the gluon distribution for the maximal momenta, we believe this implementation choice is inferior to the discretization involving the sine function.

\begin{figure}
\begin{center}
\includegraphics[width=0.5\textwidth]{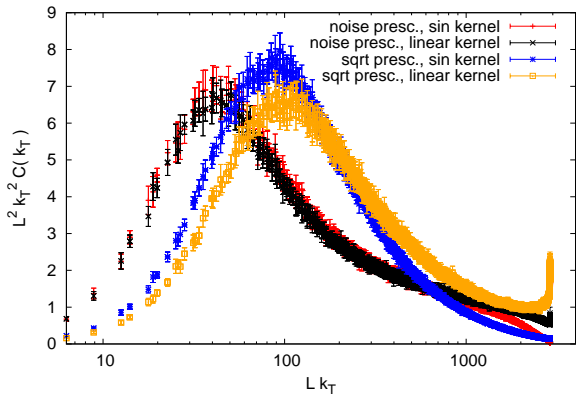}
\caption{Comparison of gluon densities obtained with different $\alpha_S$ prescriptions and different kernel discretizations at $s=0.16$ and $L/a=1024$. The linear kernel shows worse behavior close to the boundary as the distribution rises for large $k_T$. In the case of the ``square root'' prescription, the distortions of the linear kernel also propagate to small $k_T$. We conclude that the linear kernel should not be used in subsequent calculations.\label{fig. running coupling kernel}}
\end{center}
\end{figure}

Finally, we can compare the results obtained under evolution implemented in position and momentum spaces. Note, as already signaled in the introduction, that the Langevin equation is written only in position space. The Fourier acceleration used to perform some of the computations in momentum space may introduce uncontrolled discretization/finite volume effects. In Fig.~\ref{fig. running coupling comparison}, we show the comparison of evolved gluon distributions performed in both spaces at different values of the rapidity $s$. Although, as suspected, some differences can be seen in the shape of the gluon distributions, their size is negligible on the scale of the effect of the evolution itself. The discrepancies appear for the maximal transverse momenta $L k_T$ where the boundary conditions and finite volume effects are maximal. The above observation remains true for both running coupling prescriptions. Since the evolution in position space is computationally much more demanding, we can therefore conclude that from a practical perspective the solutions obtained in momentum space are reliable and can be used in subsequent phenomenological studies. However, for observables sensitive to the large $k_T$ tail, some care must be taken.

\begin{figure}
\begin{center}
\includegraphics[width=0.5\textwidth]{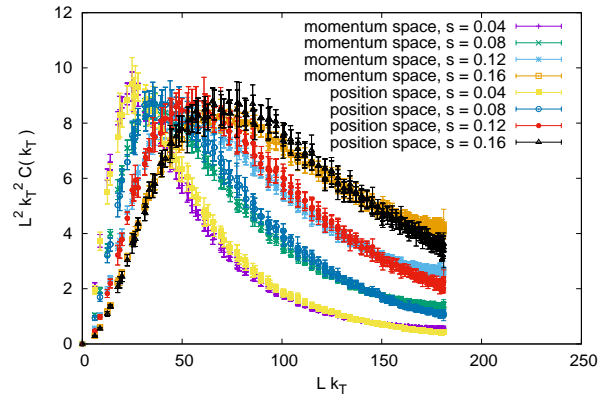}
\includegraphics[width=0.5\textwidth]{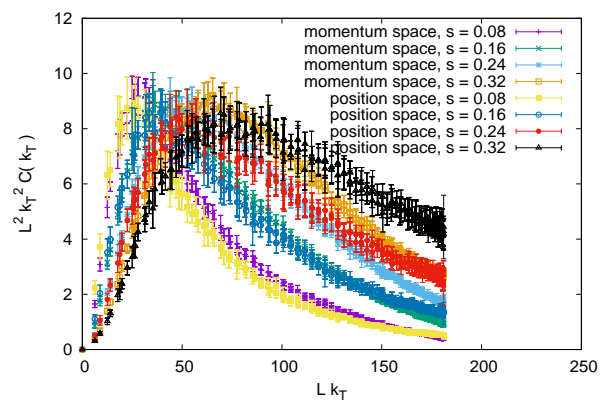}
\caption{Comparison of gluon densities obtained with evolution implemented in position and momentum spaces using the ``square root'' (top) and noise (bottom) $\alpha_s$ prescriptions on a lattice with extent $L/a=64$. \label{fig. running coupling comparison}}
\end{center}
\end{figure}

\section{Discussion}
\label{sec. conclusions}

Finally, we are in position to discuss physical implications of the different implementation choices. As our first observable, we study the evolution speed, i.e.\ the rapidity dependence of the derivative of the saturation scale. In Fig.~\ref{fig. derivative}, we plot the quantity
\begin{equation}
    \lambda = \frac{d \ln Q_s^2 L^2}{d y},
\end{equation}
where 
\begin{multline}
y = \frac{\pi^2 s}{\alpha_s(Q_s)} \approx \frac{2 \pi^2 s \beta_0 \ln \big( Q_s/\Lambda_{\textrm{QCD}} \big) }{4\pi} =\\= 
\frac{1}{2} \pi s \beta \ln \big( Q_s/\Lambda_{\textrm{QCD}} \big),
\end{multline}
where we employed the definition Eq.~\eqref{eq. alpha_s in momentum} of $\alpha_s(\mathbf{k})$. In the previous section, we have demonstrated that our results are robust against systematic uncertainties, i.e.\ increasing the volume or changing the evolution between the momentum space and position space implementations provides results compatible within their uncertainties. We demonstrate this once again for the observable $\lambda$ in the top panel of Fig.~\ref{fig. derivative} where the dependence on the lattice extent is plotted. Clear agreement of all data sets corroborates again the control of systematic effects. Conversely, the bottom panel demonstrates that changing the running coupling prescription has a genuine physical effect, as the rate of change of the saturation scale is different between the ``square root'' and ``noise'' prescriptions for small saturation scales, up to around $10\Lambda_{\textrm{QCD}}$. This is even true with our rather conservative estimates of the systematic uncertainty in the extraction of $Q_s$.
Fig.~\ref{fig. derivative} also contains another set of data points. As a consistency check, we have determined the evolution of the saturation scale with rapidity using an alternative definition of the saturation scale. We follow Ref.~\cite{Lappi:2012vw} where $Q_s$ was fixed from the gluon correlation function in position space, 
\begin{equation}
    C\left(\frac{R_s}{L} = \frac{\sqrt{2}}{Q_s L}\right) = e^{-\frac{1}{2}}.
    \label{eq. position definition saturation scale}
\end{equation}
Although this is only a change in the definition, it has a significant effect on the evolution speed $\lambda$, as can be inferred from Fig.~\ref{fig. derivative}. At small saturation scales, i.e. $Q_s$ smaller than $10-15\Lambda_{\textrm{QCD}}$, for both running coupling prescriptions the value of $\lambda$ defined from the definition Eq.~\eqref{eq. position definition saturation scale} is larger than $Q_s$ obtained from the maximum of the correlation function in momentum space. At large saturation scales, all prescriptions and definitions seem to converge to a single value of around $\lambda \sim 0.2$. At the technical level, the definition \eqref{eq. position definition saturation scale} suffers from large discretization effects for large saturation scales, which is visible as a scatter of data. This is because the exponential decay of the correlation function is loo large to precisely estimate the intercept with the $e^{-\frac{1}{2}}$ line. This effect should decrease when the lattice size is increased. Conversely, the definition in momentum space described in Section \ref{sec. systematics} is more reliable in this region, giving smaller statistical and systematic uncertainties. Again, both definitions and implementations yield quantitatively similar values for the evolution speed at large saturation scales.

\begin{figure}
\begin{center}
\includegraphics[width=0.5\textwidth]{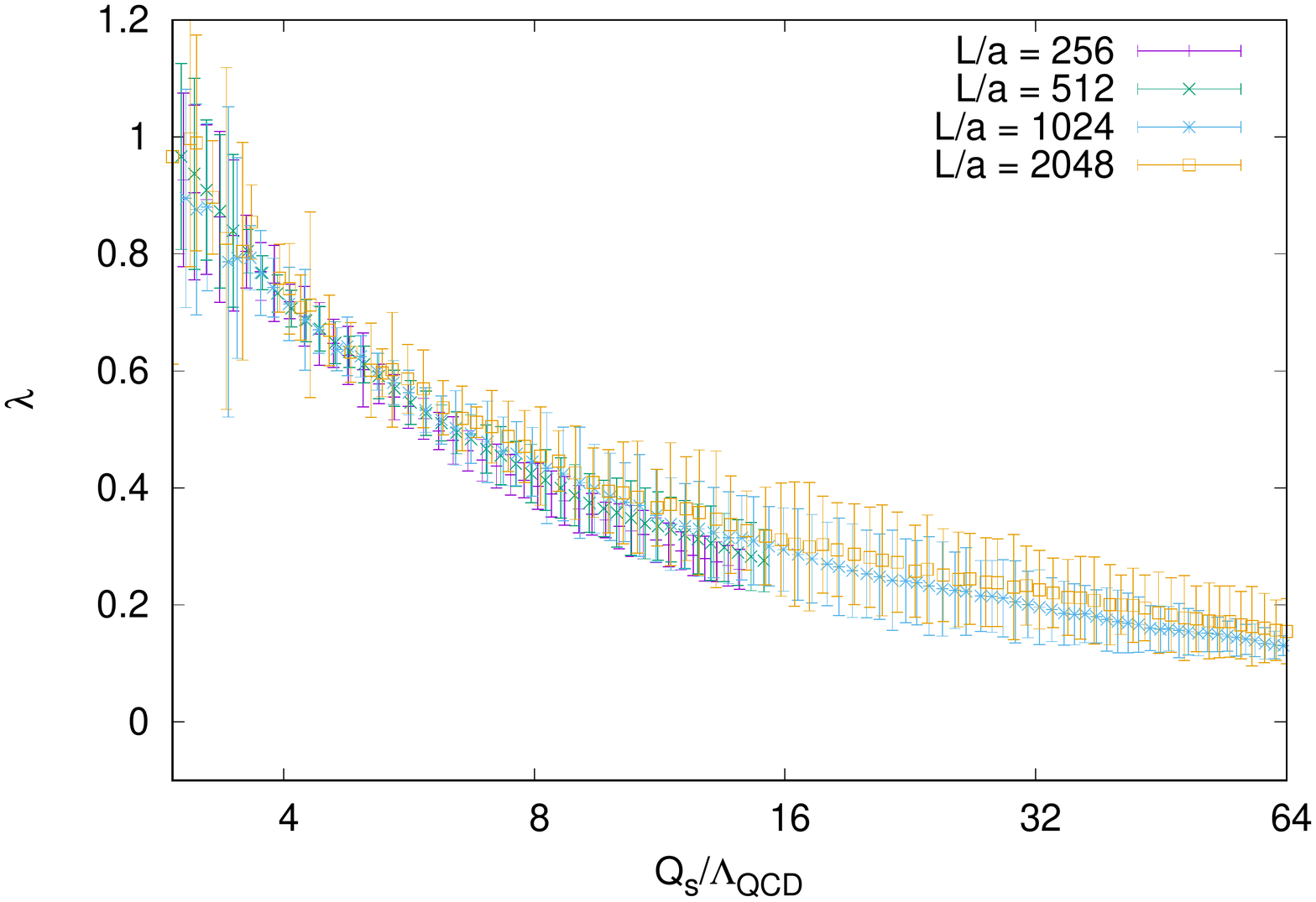}
\includegraphics[width=0.5\textwidth]{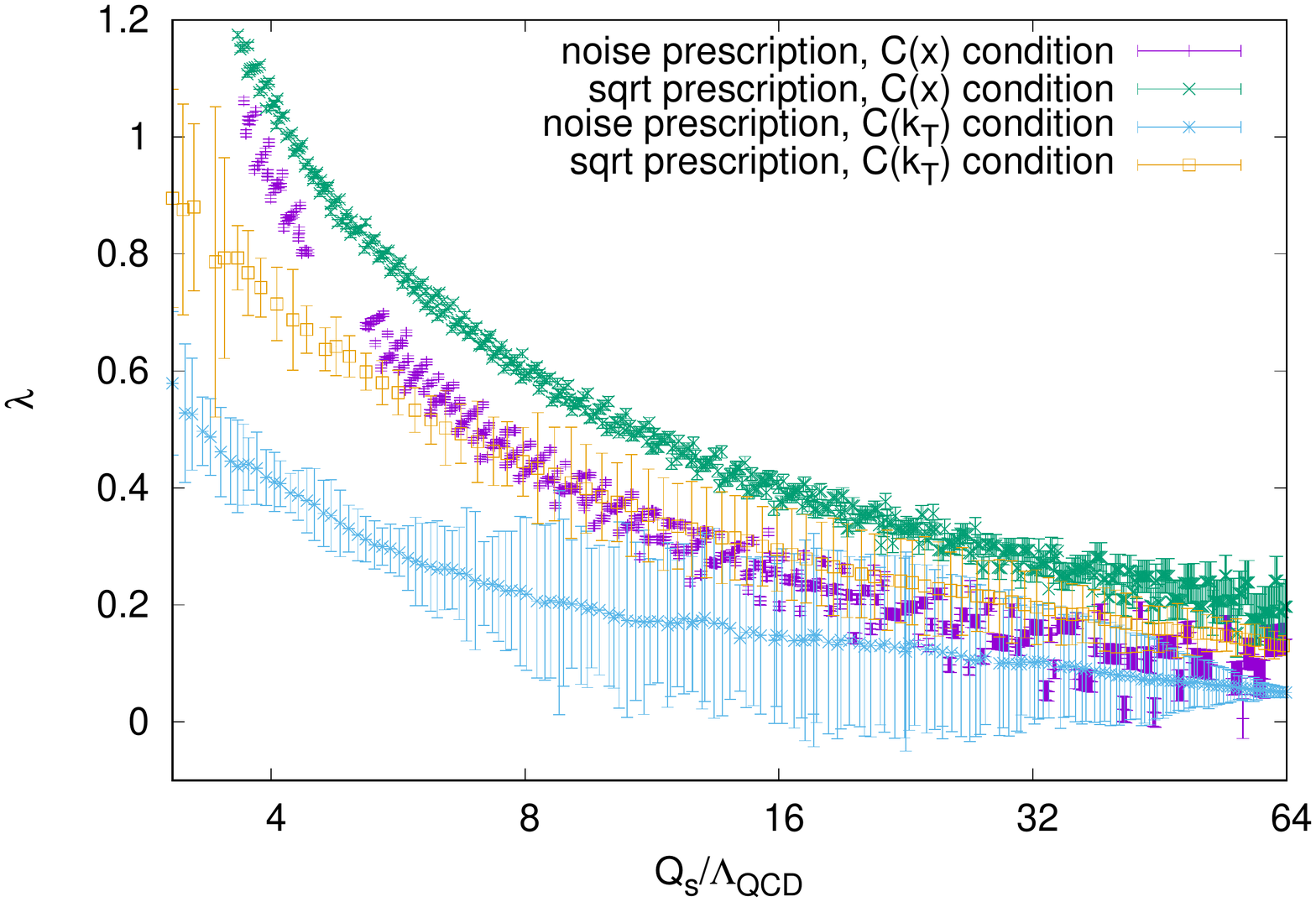}
\caption{Derivative of the saturation scale with respect to the rapidity $y$ for different definitions of $Q_s$ and running coupling prescriptions. Top: Volume dependence for the ``square root'' prescription. We check the reliability of the results by comparing $\lambda$ for increasing lattice extents. No significant deviations are seen in the range of $Q_s$ accessible on each lattice volume. Bottom: All data were simulated on a lattice with extent $L/a=1024$, except for the ``square root'' prescription and position space definition of saturation scale where a lattice of extent $L/a=2048$ was used. Also, two different definitions of the saturation scale are employed, involving gluon distributions in either momentum space or position space. \label{fig. derivative}}
\end{center}
\end{figure}

Secondly, we present for the first time the results obtained using the Hatta-Iancu running coupling prescription, see Fig.~\ref{fig. hatta}. Since the definition of the latter is provided in position space as a function of the separation in the final correlation function, we keep all the calculations in position space, avoiding Fourier transforms. In order to show the physical effect of the Hatta-Iancu prescription, we contrast the position-space correlation function obtained using this prescription with the one from the ``square root'' coupling. We observe statistically significant differences at small separations. We plot results for two volumes, $L/a=48$ and $L/a=64$, which enable us to draw the conclusion that the differences cannot be attributed to finite volume effects, but to the intrinsically different physical effect of the running coupling constant.
We note that the slower decay of the correlator in the Hatta-Iancu case implies, according to Eq.~\eqref{eq. position definition saturation scale}, smaller values of the saturation scale.

\begin{figure}
\begin{center}
\includegraphics[width=0.5\textwidth]{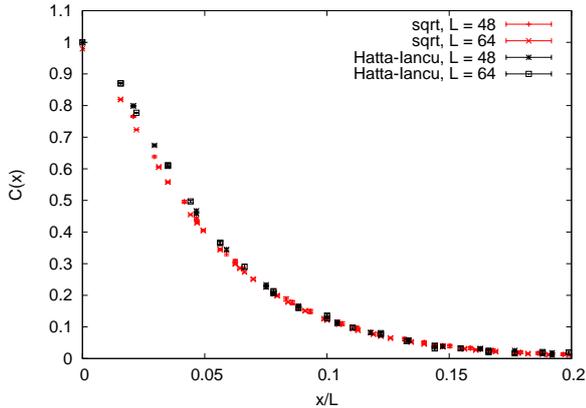}
\caption{Comparison of correlation functions in position space calculated using the ``square root'' and Hatta-Iancu prescriptions. Statistically significant deviations can be observed for small separations. Results for two volumes, $L/a=48$ and $L/a=64$ are shown, pointing to the conclusion that the differences cannot be attributed to finite volume effects but to the intrinsically different physical effect of the running coupling constant.\label{fig. hatta}}
\end{center}
\end{figure}

Finally, we discuss the large-$k_T$ asymptotics, for which analytic arguments exist in the large rapidity limit. The correlator at large rapidity is a difficult observable, as it requires large lattice extents and one has to make sure that the appropriate scaling regime is reached. It is expected that the $1/k_T^2$ behaviour \cite{Kovchegov:2012mbw}, present in the distribution corresponding to the initial condition, will be modified by the evolution to a less steep dependence. In Fig.~\ref{fig. asymptotics}, we demonstrate the situation when the effects of the running coupling constant are included for two large lattices. Two regimes can be identified: the first one for moderate values of $k_T L$, where both volumes give comparable results, and the second for large $k_T L$, where finite volume effects distort the tail of the distributions. Although the volume independence of the data in the first region may indicate that the scaling regime is reached, the resulting power law dependence strongly depends on the details of the implementation and of the running coupling constant prescription. The separation between the two regions is marked roughly by the black vertical line in the figure. Therefore, this suggests that the proposed prescriptions, not only yield values for the saturation scale which differ by a rough factor of 2, but they also differ significantly in the regime of large transverse momenta. For the data shown, the fitted power law $(L/a)^2 k_T^2  \tilde{C}(L k_T)=ax^b$ yields $b\approx-0.5$ for the ``noise'' prescription and $b\approx-1$ for the ``square root'' prescription. It is a question for further studies how the exponent in the power law depends on rapidity. As shown in Fig.\ref{fig. hatta}, the Hatta-Iancu prescription is equivalent to the ``square root'' definition at large distances, hence the differences we observe at small distances should translate to differences in the scaling for large $k_T$. The Fourier transform, needed to go from position to momentum space, requires the knowledge of the correlation function at all positions on the lattice, which is prohibitively expensive for the current numerical setup. Therefore, we do not have reliable data from large enough lattice sizes to include the large $k_T$ asymptotics with the Hatta-Iancu prescription.

\begin{figure}
\begin{center}
\includegraphics[width=0.5\textwidth]{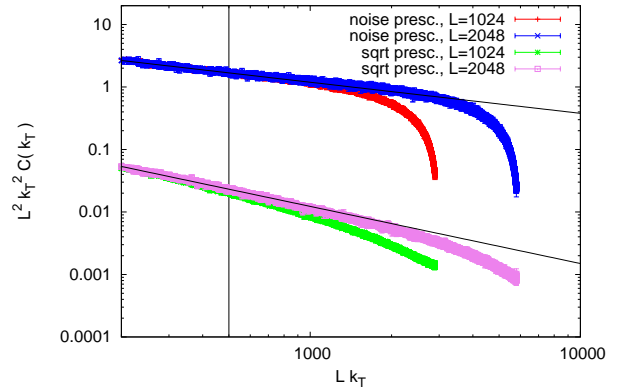}
\caption{Large $k_T L$ dependence of the evolved gluon distribution for different implementations and running coupling prescriptions at $s=0.16$. A power law, $(L/a)^2 k_T^2  \tilde{C}(L k_T)=ax^b$, is fitted to the data on the left of the vertical black line, with $a$ and $b$ being fit parameters. The fitted exponent $b$ is around $-0.5$ ($-1$) in the case of the ``noise'' (``square root'') prescription.  In order to increase the figure readability, the ``square root'' data were divided by a factor 100. \label{fig. asymptotics}}
\end{center}
\end{figure}

To summarize, we have presented a systematic comparison of solutions of the JIMWLK equation using numerical, lattice-related techniques. On the technical side, among the main systematic effects that we investigated are momentum/position space implementation of the evolution, dependence on the volume, Langevin step size and kernel discretizations. On the phenomenological side, we discussed the inclusion of the effects of the running of the strong coupling constant following three prescriptions proposed in the literature. All these results are fundamental for a reliable program of relating the numerical solution of the JIMWLK equation to experimental data (such as the $F_2$ structure function), since they give constraints on the parameters of the numerical setup such that the different systematic uncertainties are under control. Thus, they allow one to optimize the numerical setup and ensure efficient and robust fitting with quantified errors.

\section*{Acknowledgements}
P. Korcyl and K. Kutak thank Ecole Polytechnique in Palaiseau for hospitality and acknowledge fruitful discussions with Claude Roiesnel. C. Marquet is also grateful to Claude Roiesnel for earlier collaboration on the topic. We also acknowledge useful discussions with T. Lappi and email exchanges and discussions with Y. Hatta.  Computer time allocations 'nspt', 'pionda', 'tmdlangevin' and 'plgtmdlangevin2' on the Prometheus supercomputer hosted by AGH Cyfronet in Krak\'{o}w, Poland were used through the polish PLGRID consortium. P. Korcyl acknowledges partial support by the Polish National Science Center, grant No. 2017/27/B/ ST2/02755, and thanks Università degli Studi di Roma "Tor Vergata" for hospitality and Polish National Agency for Academic Exchange for the Bekker fellowship during which this work was finalized. K.\ Cichy is supported by National Science Centre (Poland) Grant SONATA BIS No. 2016/22/E/ST2/00013. P. Kotko was partially supported by the  National Science Center, Poland, grant no. 2018/31/D/ST2/02731. This project has received funding from the European Union’s Horizon 2020 research and innovation programme under grant agreement STRONG – 2020 - No 824093.

\bibliographystyle{spphys}
\bibliography{references2}

 \appendix

\section{Random vectors in position and momentum spaces}
\label{sec. appendix}

We demonstrate in a discretized setup the fact that uncorrelated random vectors in position space are also uncorrelated in momentum space. We adopt the following conventions, (with $V = (L/a)^2$ )
\begin{align}
    f(k) &= \frac{1}{\sqrt{V}} \sum_n f(x) e^{-i k n} \\
    f(n) &= \frac{1}{\sqrt{V}} \sum_k f(k) e^{i k n} 
\end{align}
and
\begin{align}
    \delta(k - p) &= \frac{1}{V} \sum_n e^{i (k-p) n} \\
    \delta(n-m) &= \frac{1}{V} \sum_k e^{i k (n - m)},
\end{align}
and simplify the indices leaving only space/momentum dependence. We have as a starting point
\begin{equation}
\langle \xi_{\mathbf{x}} \xi_{\mathbf{y}} \rangle = \delta(\mathbf{x} - \mathbf{y}) = \frac{1}{V} \sum_{\mathbf{k}} e^{i \mathbf{k} (\mathbf{n} - \mathbf{m})}
\end{equation}
For the fields themselves we have
\begin{equation}
    \xi_{\mathbf{n}} = \frac{1}{\sqrt{V}} \sum_{\mathbf{k}} e^{i \mathbf{k} \mathbf{n}} \xi_{\mathbf{k}}
\end{equation}
and
\begin{equation}
    \xi_{\mathbf{k}} = \frac{1}{\sqrt{V}} \sum_{\mathbf{n}} e^{-i \mathbf{k} \mathbf{n}} \xi_{\mathbf{n}}
\end{equation}
Then we have
\begin{align}
\langle \xi_{\mathbf{p}} \xi^{\dagger}_{\mathbf{q}} \rangle &= 
\frac{1}{\sqrt{V}} \sum_{\mathbf{n}} e^{-i \mathbf{p} \mathbf{n}} 
\frac{1}{\sqrt{V}} \sum_{\mathbf{m}} e^{i \mathbf{q} \mathbf{m}}  \langle \xi_{\mathbf{n}} \xi^{\dagger}_{\mathbf{m}} \rangle \\
&= 
\frac{1}{V} \sum_{\mathbf{n}} e^{-i \mathbf{p} \mathbf{n}} 
\sum_{\mathbf{m}} e^{i \mathbf{q} \mathbf{m}}  \delta(\mathbf{n} - \mathbf{m})  \\
&= 
\frac{1}{V} \sum_{\mathbf{n}} e^{-i (\mathbf{p} - \mathbf{q})\mathbf{n}} \\
&= 
\delta(\mathbf{p} - \mathbf{q})
\label{eq dirac delta}
\end{align}
This hints to the following equivalence:
\begin{itemize}
    \item $\xi_{\mathbf{p}}$ generated in momentum space: i.e. at each lattice site we generate uncorrelated random numbers with a unit standard deviation
    \item $\xi_{\mathbf{x}}$ generated in position space: i.e. at each lattice site we generate uncorrelated random numbers with a unit standard deviation  by taking the forward Fourier transform to obtain $\xi'_{\mathbf{p}}$, it will be indistinguishable from the $\xi_{\mathbf{p}}$.
\end{itemize}
We checked numerically that the above is indeed true.
The practical choice to make is either to:
\begin{itemize}
\item generate the random vectors in one space and Fourier-transform them to the other one, or
\item generate separate random vectors in both spaces.
\end{itemize}
Obviously, both choices are equivalent in the limit of an infinite number of stochastic realizations.

\end{document}